\documentclass[12pt]{article}

\usepackage{graphicx}

\usepackage{times}
\usepackage{amsmath, amssymb}
\usepackage{newtxmath}
\usepackage{scicite}
\usepackage{url}

\newcommand{\lya}{Ly$\alpha$}

\newcommand{\mrperp}{R_\perp}

\def\cm#1{\, {\rm cm^{#1}}}
\newcommand{\mzfrb}{z_{\rm FRB}}

\newcommand{\mmstar}{M_\star}

\newcommand{\mmsun}{{\rm M}_\odot}
\newcommand{\msun}{$\mmsun$}
\newcommand{\mmhalo}{M_{\rm halo}}
\newcommand{\mhalo}{$\mmhalo$}

\newcommand{\rafrb}{21h49m23.630s}   
\newcommand{\decfrb}{-52d58m15.39s} 
\newcommand{\paell}{120} 
\newcommand{\astat}{0.555''}
\newcommand{\bstat}{0.153''}
\newcommand{\asys}{3.2''}
\newcommand{\bsys}{0.8''}
\newcommand{\zfrb}{0.47550}

\newcommand{\rmvalue}{10.9}  
\newcommand{\rmsig}{0.9}  
\newcommand{\rmunits}{{\rm rad \, m^{-2}}}
\newcommand{\mbmax}{B_\parallel^{\rm max}}   %
\newcommand{\bmax}{$\mbmax$}

\newcommand{\mdtpulse}{\tau_{\rm scatt}}
\newcommand{\dtpulse}{$\mdtpulse$}

\newcommand{\avgne}{\langle  n_e \rangle}

\newcommand{\host}{DES\,J214923.66$-$525815.28}
\newcommand{\hostsfr}{0.4}       
\newcommand{\hostsigsfr}{0.6}    %
\newcommand{\hostmstar}{9.4}     
\newcommand{\hostsigmstar}{0.19}
\newcommand{\hostebv}{0.15}      
\newcommand{\hostsigebv}{0.08}      
\newcommand{\hostsfrbeta}{0.6}   
\newcommand{\npatch}{100}   

\newcommand{\zfg}{0.3674}
\newcommand{\fgdes}{DES\,J214923.89$-$525810.43}
\newcommand{\fgname}{FG-181112} 
\newcommand{\rperpv}{29}    
\newcommand{\logmstar}{10.69}  
\newcommand{\usiglogm}{0.22}    
\newcommand{\lsiglogm}{0.46}    
\newcommand{\fgcebv}{0.43 \pm 0.19}    
\newcommand{\fgcmetal}{1.94 \, Z_\odot} 
\newcommand{\tage}{1.4}        
\newcommand{\mhalov}{12.3}    
\newcommand{\fgebv}{0.07 \pm 0.2}       
\newcommand{\fgmetal}{1.62 \, {\rm Z}_\odot}  
\newcommand{\fgsfr}{0.27}    

\newcommand{\mdmunits}{{\rm pc \, cm^{-3}}} 

\newcommand{\dmval}{589.27}   
\newcommand{\dmerr}{0.03}   
\newcommand{\dmfg}{122}   

\newcommand{\mdmfrb}{{\rm DM}_{\rm FRB}}
\newcommand{\dmfrb}{$\mdmfrb$}

\newcommand{\figfrb}{1}      
\newcommand{\figrm}{2}       
\newcommand{\figsumm}{3}     

\newcommand{\sfighostcigale}{S1}  
\newcommand{\sfighostppxf}{S2}    
\newcommand{\sfighosts}{S3}       
\newcommand{\sfigrandom}{S4}      
\newcommand{\sfigfgcigale}{S5}    
\newcommand{\sfigfgspec}{S6}     
\newcommand{\sfigbpt}{S7}     
\newcommand{\sfigsersic}{S8}    
\newcommand{\sfigDMpdf}{S9}  
\newcommand{\sfigDMcum}{S10}  
\newcommand{\sfigFB}{S11}  
\newcommand{\sfigB}{S12}  

\newcommand{\tabsumm}{1}

\newcommand{\stabphotom}{S3}
\newcommand{\stabhost}{S4}
\newcommand{\stabfg}{S5}
\newcommand{\stabflux}{S6}
\newcommand{\stabsersic}{S7}

\def\aap{Astron.\ Astrophys.}                
\def\aaps{Astron.\ Astrophys.\ Suppl.\ Ser.}                

\def\aap{Astron. \& Astrophys.}                
\newcommand{\aj}{Astron. J.}
\newcommand{\apjs}{Astrophys. J.}
\newcommand{\apj}{Astrophys. J.}
\newcommand{\apjl}{Astrophys. J.}
\newcommand{\mnras}{Mon. Not. R. Astro. Soc.}

\newcommand{\nat}{Nature}

\newcommand{\pasp}{Pub. Astron. Soc. Pacific}

\newcommand{\araa}{Annu. Rev. Astron. Astrophys.}

\newenvironment{sciabstract}{%
\begin{quote} \bf}
{\end{quote}}

\title{The low density and magnetization of a massive galaxy halo exposed by a fast radio burst}

\author
{J. Xavier Prochaska$^{1,2}$, 
Jean-Pierre	Macquart$^3$,
Matthew 	McQuinn$^4$, \\
Sunil	Simha$^1$, 
Ryan  M. Shannon$^5$,
Cherie K.	Day$^{5,6}$, \\
Lachlan	Marnoch$^{6,7}$, 
Stuart	Ryder$^7$,
Adam 	Deller$^5$,
Keith W. Bannister$^6$, \\
Shivani	Bhandari$^6$, 
Rongmon	Bordoloi$^8$,
John 	Bunton$^6$, \\
Hyerin	Cho$^9$, 
Chris	Flynn$^5$,
Elizabeth K. Mahony$^6$, \\
Chris	Phillips$^6$, 
Hao	Qiu$^{10}$, 
Nicolas	Tejos$^{11}$
\\
}

\date{}

\begin{document}



\maketitle

\noindent
1:  University of California Observatories-Lick Observatory, University of California, 1156 High Street, Santa Cruz, CA 95064, USA\\
2: Kavli Institute for the Physics and Mathematics of the Universe,
5-1-5 Kashiwanoha, Kashiwa, 277-8583, Japan\\
3:International Centre for Radio Astronomy Research, Curtin University, Bentley WA 6102, Australia\\
4:  Astronomy Department, University of Washington, Seattle, WA 98195, USA \\
5: Centre for Astrophysics and Supercomputing, Swinburne University of Technology, Hawthorn VIC 3122, Australia\\
6: Commonwealth Science and Industrial Research Organisation,  Australia Telescope National Facility, P.O. Box 76, Epping, NSW 1710 Australia \\
7: Department of Physics and Astronomy, Macquarie University, NSW 2109, Australia\\
8: North Carolina State University, Department of Physics, Raleigh, NC 27695-8202, USA \\
9: School of Physics and Chemistry, Gwangju Institute of Science and Technology, Gwangju, 61005, Korea \\
10: Sydney Institute for Astronomy, School of Physics, University of Sydney, Sydney, NSW 2006, Australia\\
11:  Instituto de F\'isica, Pontificia Universidad Cat\'olica de Valpara\'iso, Casilla 4059, Valpara\'iso, Chile\\
\normalsize{$^\ast$To whom correspondence should be addressed; E-mail:  xavier@ucolick.org.}

\baselineskip 24pt

\newpage 
\begin{sciabstract}
Present-day galaxies are surrounded by cool and enriched halo gas 
extending to hundreds of kiloparsecs. 
This halo gas is thought to be the dominant reservoir of material available to fuel future 
star formation, but 
direct constraints on its mass and physical properties have 
been difficult to obtain. 
We report the detection of a fast radio burst 
(FRB 181112) with arcsecond precision, which passes
through the halo of a foreground galaxy. 
Analysis of the burst shows the halo gas has low net magnetization and turbulence. 
Our results imply 
predominantly diffuse gas in massive galactic halos, even those 
hosting active supermassive black holes, contrary to some
previous results.

\end{sciabstract}


The low-density gas located in the outskirts of galaxies influences
the process of galaxy formation, 
especially gas accretion and feedback
\cite{tpw17}.
Absorption-line spectroscopy can detect
this nearly invisible medium. 
Surveys
demonstrate a very high incidence 
of cool gas (with
temperature $T \sim 10^4$\,K), detected through hydrogen
Lyman 
series and continuum absorption, surrounding galaxies with
masses similar to our Milky Way (e.g.\ \cite{pwc+11,tpw17}). 
Properties of this gas depend on galaxy mass but are otherwise
insensitive to the galaxy's internal properties
\cite{thom12,tpw17,p+17,bordoloi18}.
Estimates for the total mass of the cool gas 
match or exceed the baryonic mass of the galaxy \cite{werk+14,p+17}.
Theoretical treatments 
of halo gas around present-day galaxies disagree on the
proportion of total mass retained in the halo
during galaxy formation, with 
estimates ranging from several tens of percent to all of the
baryons predicted to accrete into the halo
(e.g.\ \cite{hafen+18,pillepich+18}).
This uncertainty stems from observational insensitivity to the hot ($T \gtrsim 10^6$\,K) gas which pervades galaxy 
halos (and within which the cold gas is embedded), 
and from systematic uncertainties in estimating its mass \cite{werk+14,tpw17}.  
Constraints on the density and temperature of the halo gas are 
sufficiently limited to
allow qualitatively different descriptions of its ionization 
and distribution \cite{faerman17,stern+18}.
The origin of the cool gas and its composition are 
challenging to explain theoretically; 
some models require cosmic rays and magnetic fields 
to transport material from the central galaxy to sustain the 
cool medium (e.g.\ \cite{ji18}).

%
At Coordinated Universial Time 17:31:15.48365 on 2018 Nov 12 the Commensal Real-time ASKAP Fast Transients (CRAFT)
survey on the Australian Square Kilometer Array
Pathfinder (ASKAP) detected a fast radio burst (FRB~181112) 
from the $12$ antennas observing at the time.
The burst arrival time swept across the observing band ($\approx 1.129-1.465$\,GHz, Fig.~{\figfrb}A)
due to propagation of the burst through the
foreground plasma. 
The burst sweep yields an estimate of the FRB dispersion measure 
$\mdmfrb = \dmval \pm \dmerr~\mdmunits$ which is
the integrated density of electrons 
$n_e$ at distance $r$ from Earth
scaled by $(1+z)^{-1}$ with $z$ the redshift:
$\mdmfrb \equiv \int n_e/(1+z) \, dr$.
The real-time detection triggered full download of the voltage data; 
these precisely localized the burst to a sky position 
${\rm \rafrb}, \, {\rm \decfrb}$ 
(right ascension, declination, J2000) 
with a statistical
(systematic) error ellipse oriented at \paell\,deg E of N on
the sky with major axis $a = \astat (\asys)$ and 
minor axis
$b = \bstat (\bsys)$
\cite{supp}. 

Fig.~{\figfrb}B shows a $g$-band image centered
on FRB~181112 obtained with the 
FOcal Reducer/low dispersion Spectrograph 2 (FORS2) 
instrument on the Very Large Telescope (VLT). 
It shows the presence
of a galaxy coincident with FRB~181112, previously cataloged by
the Dark Energy Survey (DES; \cite{desdr1}) as \host.  
The DES and FORS2 data 
also show a luminous galaxy 
$\approx 5''$ to the North of the FRB event (\fgdes). 
We used follow-up spectroscopy 
with the FORS2 instrument 
to measure the redshift \cite{supp} of the former
galaxy as $z = \zfrb$, 
and the latter galaxy
as $z = \zfg$, i.e.\ in the foreground.
We associate FRB~181112 with \host. 
Compared to the other three known host galaxies of FRBs,
the host galaxy of FRB~181112
has an intermediate stellar mass of
$\mmstar \approx 10^{\hostmstar}$ solar masses (\msun) (Fig.~\sfighosts;
\cite{supp}).  
It has colors matching star-forming galaxies at $z \sim 0.4$,
has an estimated star formation rate of 0.6\,\msun\,yr$^{-1}$,
and shows no signatures of an active galactic nucleus (AGN)
\cite{supp}.

The FRB sightline 
passes at an impact parameter $\mrperp = \rperpv$\,kpc
from \fgdes\ (hereafter referred to as \fgname)
allowing us to probe the halo of this foreground galaxy.
We analyzed the DES, FORS2, and complementary longer-wavelength
Wide-field Infrared Survey Explorer (WISE)
data 
to determine
\fgname's 
physical properties \cite{supp}, 
We derive
a high stellar mass $\log_{10} \mmstar/\mmsun = 
\logmstar^{+\usiglogm}_{-\lsiglogm}$,
nebular emission lines indicative of an AGN
and classifying it as a Seyfert galaxy, and
an old ($> \tage$\,Gyr) quiescent stellar population
(Tables~\tabsumm\ and \stabfg).
Surveys of the halo gas surrounding galaxies of
similar mass,
with or without AGN activity \cite{berg+18},
almost ubiquitously reveal
strong cool ($T \sim 10^4$\,K)
gas absorption for sightlines $\mrperp \le 100$\,kpc.
Generally,
the inferred total column densities of ionized gas 
exceed $10^{20} \cm{-2}$ \cite{werk+14,p+17}, and
transitions of heavy elements indicate a turbulent
velocity field \cite{cwt+10} suggesting that
a fraction of the gas has a
relatively high density ($n_{\rm H} \sim 1 \cm{-3}$; \cite{lm18}).
Such a foreground medium should impact the FRB signal.

The column of gas close to this massive galaxy, however,
does not dominate \dmfrb. It contributes only
DM$_{\rm FG} \sim 50-120~\mdmunits$, depending on 
assumptions for the density profile and total mass
of the halo gas \cite{supp}.
The measured \dmfrb\ is consistent with
models that include cosmic gas, our Galaxy, and the host
(Fig.~\sfigDMpdf; \cite{mcquinn14,xyz19}).
The sightline to FRB~181112 also intersects the edge
of the Fermi Bubbles \cite{supp},  a complex of hot gas encompassing 
the Galactic center. 
The expected DM contribution from gas in these bubbles is 
small \cite{supp}, but their entrained magnetic
field may contribute to the FRB rotation measure. 

The rotation measure
RM 
is the density-weighted
integral of the magnetic field parallel to the FRB sightline.
The voltages recorded from the ASKAP antennas measure the 
electric field at the antenna locations in two orthogonal directions on the plane 
of the sky, enabling the linear polarization fraction of the burst radiation (and its position angle) to be measured as a function of frequency.
Averaged over its duration we find the burst to be approximately 90\% linearly polarized and 10\% circularly polarized \cite{supp}.
This can be used to estimate the burst RM, as PA$_{\rm obs}$~$=$ PA$_{\rm int} + (c/\nu)^2$RM, where $\nu$ is the frequency
and PA$_{\rm obs}$ and PA$_{\rm int}$ are the observed and intrinsic polarization angles, respectively.
Fig.~\figrm\ depicts the frequency sweep of the polarization
angle; the apparent $\nu^{-2}$ frequency dependence is the
RM signature.  We fitted an RM to the sweep, yielding
RM~$= \rmvalue \pm \rmsig \, \rmunits$.
This is a low RM value, consistent (within the uncertainty) 
to the estimated RM due to our Galaxy
towards FRB\,181112 \cite{supp}.
Adopting an upper limit of RM$ \, < 11 \, \rmunits$, we calculate an upper limit 
for the maximum parallel magnetic field
\bmax\ in the halo of \fgname, 
$\mbmax < 0.8 {\muup \rm G} \, (n_e/10^{-3} \, {\rm cm^{-3}})^{-1} 
\, (\Delta L/30\,{\rm kpc})^{-1}$ 
in the limit of a perfectly ordered magnetic field
with $\Delta L$ a characteristic length-scale through the halo. 
We have adopted fiducial values for $n_e$ and $L$
that may characterize the halo of \fgname\ (similar to those
adopted for the DM$_{\rm FG}$ estimation).
Field reversals would lead us to underestimate \bmax.
Nevertheless, this low value for \bmax\
implies that either the magnetic field in the halo is low compared to the interstellar medium, 
or that it is largely disordered.

These constraints have implications for the circumgalactic gas.  The magnetic field value in equipartition with the thermal energy of the virialized halo gas is
$B_{\rm eq} \equiv \sqrt{8 \pi n_e k_b T} = 2 \upmu {\rm G} (n_e/10^{-3} \, {\rm cm^{-3}})^{1/2} (T/10^6 {\rm K})^{1/2}$ with
$k_b$ the Boltzman constant. Our 
\bmax\ limit is comparable to $B_{\rm eq}$ for physically motivated $n_e$, $\Delta L$ and $T$, constraining the magnetic field to be near or below equipartition if the total field is comparable to the net parallel field.
Magnetic fields around the equipartition value enhance the rate of condensation of the hot circumgalactic gas into cooler clouds \cite{ji18} as well 
as the survival of cool accreting gas  \cite{berlok19}.  Near equipartition field strengths are generated in some models in which cosmic ray pressure transports cool gas and metals to large distances from galaxies \cite{pakmor2016,butsky2018}. 
Our limit on \bmax\ is below the mean estimate 
for sightlines that show strong gas absorption \cite{bernet08} despite
our sightline likely intersecting gas with 
similarly strong absorption in \fgname\ (e.g.\ \cite{thom12}).  

The halo gas of \fgname\ broadens the width of the
pulse at any given frequency.
This temporal broadening \dtpulse\ arises from density fluctuations within the
medium which impose small light-travel differences for 
rays propagating through the gas \cite{macquart13,mcquinn14}.
This scattering is geometrical and its effects are 
maximal for a scattering `screen' located at one-half the
distance to the FRB.
%
We determine an
upper limit of $\mdtpulse < 40\,\upmu$s due 
to scattering,
constraining both the turbulent properties of the halo gas and its density.   
A pulse with 150 times higher width (3\,ms) would have still been detected, i.e.\ the very narrow
width of FRB~181112 is not the result of observational bias.
Fig.~{\figrm}B shows that the temporal profile of FRB~181112 consists of two pulses separated by approximately 800~$\upmu$s.  The broadening limit is derived by modeling each component as a symmetric intrinsic pulse convolved with the one-sided exponential decay expected due to scattering (see \cite{supp}). Temporal smearing due to inhomogeneities in the plasma distribution along the line of sight would otherwise broaden the pulse to a frequency-dependent duration $\mdtpulse(\nu) = \tau_0 (\nu/{1 \, \rm GHz})^\gamma$, where the index $\gamma$  is typically $\approx -4$ 
\cite{supp}.

The observed \dtpulse\ constrains the integral of 
the square of the density along the sightline, 
$\int dx \delta n_e(x)^2$, which we relate to the electron column density with the parameterization
$\avgne \Delta L^{1/2} = \alpha^{-1}  \left (\int  dx   \delta n_e^2 \right)^{1/2}$, which takes the halo of \fgname\ to have characteristic length $\Delta L$ with an average density of $\avgne$. Thus, the parameter $\alpha$ encapsulates the root-mean-square 
amplitude of density fluctuations and the volume filling fraction of the turbulence, $f_V$.    The limit on the in situ density assuming a Kolmogorov spectrum of turbulence \cite{supp} is 
\begin{equation}
\langle n_e \rangle < 2 \times 10^{-3} \alpha^{-1}  \left( \frac{\Delta L}{50\,{\rm kpc}} \right)^{-1/2} \left( \frac{L_0}{1\,{\rm kpc}} \right)^{1/3}  \left(\frac{\mdtpulse}{40 \, \rm \upmu s} \right)^{5/12}  
\,{\rm cm}^{-3},
\label{eqn:scatteringconstraint}
\end{equation}  
where $\Delta L \sim 50$\,kpc approximates the path length through the foreground halo and $L_0$
is the outer scale of turbulence.  
As the turbulence is likely to be sourced by galactic winds and inflows, we expect it to be driven at 
scales less than the impact parameter ($\sim 30$\,kpc)
and consider $L_0 = 1$\,kpc a reasonable value.

We now examine two standard
models for halo gas in which the medium is
comprised of either hot ($T \sim 10^6\,$K)
virialized gas or cool gas 
pressure-confined by the hot gas.
In the case of hot virialized gas,  our constraint 
on $\avgne$ suggests densities lower than those 
expected of $\sim 10^{-3}$~cm$^{-3}$ gas with 
kiloparsec driving scales (see Fig.~\sfigB).  
Because we expect the volume filling factor of this gas to be near unity, the upper limit on the density can only be ameliorated if the gas is much less turbulent (i.e. $\alpha \ll 1$) relative to galactic astrophysical plasmas, especially the interstellar medium of our Galaxy, where $\alpha \sim 7$ \cite{AnanNara1988, supp}.  

For turbulent, cool $10^4$K clouds  embedded in a hot medium, the constraints are stronger. 
Assuming pressure equilibrium 
with characteristic values for the hot gas $n_e = 10^{-3}$~cm$^{-3}$ and $T=2\times 10^6~$K, application of equation~(\ref{eqn:scatteringconstraint}) with $L_0=1~$kpc and $\Delta L=50~$kpc
yields $\alpha < 0.01$. Since $\alpha \propto f_V^{1/2}$, we require a filling factor of cool clouds of
$f_V < 10^{-4}$ if the clouds are fully turbulent. Even lower values are required to satisfy this condition if
the driving for turbulence within cool clouds 
is instead at parsec scales, which may be physically motivated \cite{mccourt+18}.  

These limits on the halo gas density derived from the scattering analysis contradict prior inferences that cool halo gas has a volume filling fraction of $f_V \sim 10^{-3} - 10^{-2}$ \cite{cantalupo14, werk+14, lan17}. The total neutral hydrogen column density
offers the most direct comparison to our result: 
photoionization equilibrium constrains the same combination of parameters as scattering, implying $\left(n_e/ 0.1~{\rm cm}^{-3}\right)$ $\left( f_{V}/ 10^{-3} \right)^{1/2}$ \,$\left( \Delta L/ 50\,{\rm kpc} \right)^{1/2} \sim 1$ if we take a typical neutral hydrogen
column density of $10^{18}~$cm$^{-2}$ at $30~$kpc measured for halos with similar masses as \fgname\ \cite{p+17}.  
Reconciling these values with the scattering from \fgname\ either implies that the cool clouds are less turbulent than assumed or that our sight line has less cool gas than is typical.  
 The foreground galaxy is classified as a Seyfert, with an embedded accreting supermassive black hole in a central AGN that 
 could lead to a more evacuated halo \cite{choi+14}, although it has been argued that such activity may lead to more cool gas \cite{david+14}.   
 Even if the clouds are not turbulent and instead we consider the refractive bending of light through a network of parsec-scale clouds \cite{mccourt+18}, we rule out a population 
 of $0.1~$pc clouds or smaller with $f_V \sim 10^{-3}$ \cite{supp}.


FRBs experience a number of propagation effects
which render them sensitive probes of the density, magnetic field and turbulence of the otherwise elusive gas that pervades galaxy halos.  
The constraints derived from FRB\,181112
for the halo of a massive galaxy are summarized 
in Fig.~\figsumm.
The $n_e, B_\parallel$ parameter space ruled out
by our observations conflicts
with several previous inferences for halo gas
\cite{bernet08,cantalupo14,lan17}.
Our observations 
indicate a density of hot gas that is lower than in many 
models and also a column of cool gas that is smaller than commonly inferred.  

These results establish intergalactic optics as a potent means of elucidating the physical properties of the diffuse gas in the halos of galaxies.  It may further be that the multiple pulses observed in FRB 181112 are due to multipath propagation through the gas. This would be a natural consequence of a medium comprising very low filling-factor cool clouds embedded in hot virialized halo gas, with the pulse multiplicity signifying the number of clouds intersected, and their arrival times yielding their offsets from the direct burst sight line.  This work represents only the first step in harnessing the impulsive properties of FRBs to resolve the structures of galaxy halos.






\section*{Acknowledgments}
Based on observations collected at the European Southern Observatory under ESO programme 0102.A-0450(A).
We are grateful to Australia Telescope National Facility (ATNF) operations staff and the Murchison Radio-astronomy Observatory staff for supporting our ASKAP operations, and the ATNF Director and Steering Committee for dedicating time for these observations. 
K.W.B., J.P.M, and R.M.S. acknowledge Australian Research Council (ARC) grant DP180100857.
 A.T.D. is the recipient of an ARC Future Fellowship (FT150100415).
R.M.S. acknowledges support through ARC grants and CE170100004. N.T. acknowledges
support from PUCV research funding 039.333/2018.
Work at the Naval Research Laboratory is supported by NASA.
The Australian Square Kilometre Array Pathfinder, Australia Telescope Compact Array, and Parkes Radio Telescope are part of the Australia Telescope National Facility which is managed by CSIRO. 
Operation of ASKAP is funded by the Australian Government with support from the National Collaborative Research Infrastructure Strategy. ASKAP uses the resources of the Pawsey Supercomputing Centre. Establishment of ASKAP, the Murchison Radio-astronomy Observatory and the Pawsey Supercomputing Centre are initiatives of the Australian Government, with support from the Government of Western Australia and the Science and Industry Endowment Fund. 
We acknowledge the Wajarri Yamatji as the traditional owners of the Murchison Radio-astronomy Observatory site. 

\subsection*{Funding}
 J.X.P. and S.S. are supported by the National Science
 Foundation grant AST-1911140.
 K.W.B., J.P.M, and R.M.S. acknowledge Australian Research Council (ARC) grant DP180100857.
 A.T.D. is the recipient of an ARC Future Fellowship (FT150100415).
 R.M.S. acknowledge support through ARC grant FL150100148 and 
  through ARC grant CE170100004. 
 N.T. acknowledges support from PUCV research funding 039.333/2018.

\subsection*{Author contributions}
J.X.P., J.P.M., M.M., S.S., R.M.S, C.F., and C.K.D drafted the manuscript.
J.X.P., N.T., S.R., L.M, S.S., and E.M. obtained, reduced and interpreted optical observations.  R.B. did the {\it FERMI} bubble analysis.
K.W.B. built the search and voltage capture software. A.D., C.P., C.K.D. H.C.,H.Q. and S.B., designed, built and conducted the correlation, calibration and imaging software to localize FRB~181112. R.M.S. led the ASKAP observing, and interpreted radio band polarization data. 
E.M. and S.B. and obtained, reduced, and interpreted ATCA data. J.B. designed the ASKAP digital systems.

\subsection*{Competing Interests}
The authors declare no competing interests.

\subsection*{Data and materials availability:}

Based on observations collected at the European Southern Observatory under ESO programs 0102.A-0450(A) and 2102.A-5005(A) (PI: Macquart), available from \url{http://archive.eso.org/}. 
Observations from  the Australia Telescope Compact Array under program C3211 (PI: Shannon), and the Parkes radio telescope under program P958 (PI: Shannon) can be retrieved from \url{https://atoa.atnf.csiro.au/}. 

Further datasets used in this paper are available from the gSTAR Data Management and Collaboration Platform (gDMCP) at \url{https://data-portal.hpc.swin.edu.au/} using the tag FRB181112 : 
nine visibility data sets used to calibrate and determine the localisation of FRB~181112, radio images of the FRB and the surrounding field, and the ATCA images used for astrometric alignment.

Data reduction scripts and code written by the co-authors for this project are available from the {\sc craft} git repository \url{https://bitbucket.csiro.au/scm/craf/craft.git}, the {\sc psrvlbireduce} repository \url{https://github.com/dingswin/psrvlbireduce}, 
the {\sc FRB} repository
\url{https://github.com/FRBs/FRB},
and the {\sc PypeIt} repository
\url{https://github.com/pypeit/pypeit}.

\subsection*{Supplementary Materials:}

{\bf This PDF file includes}\\
Materials and Methods\\
Supplementary Text\\
Figures S1 to S12\\
Tables S1 to S7 \\
References \textit{(30-84)}

\clearpage

\begin{figure}[h]
\begin{center}
\vskip -1.5in
\includegraphics[width=1.1\linewidth]{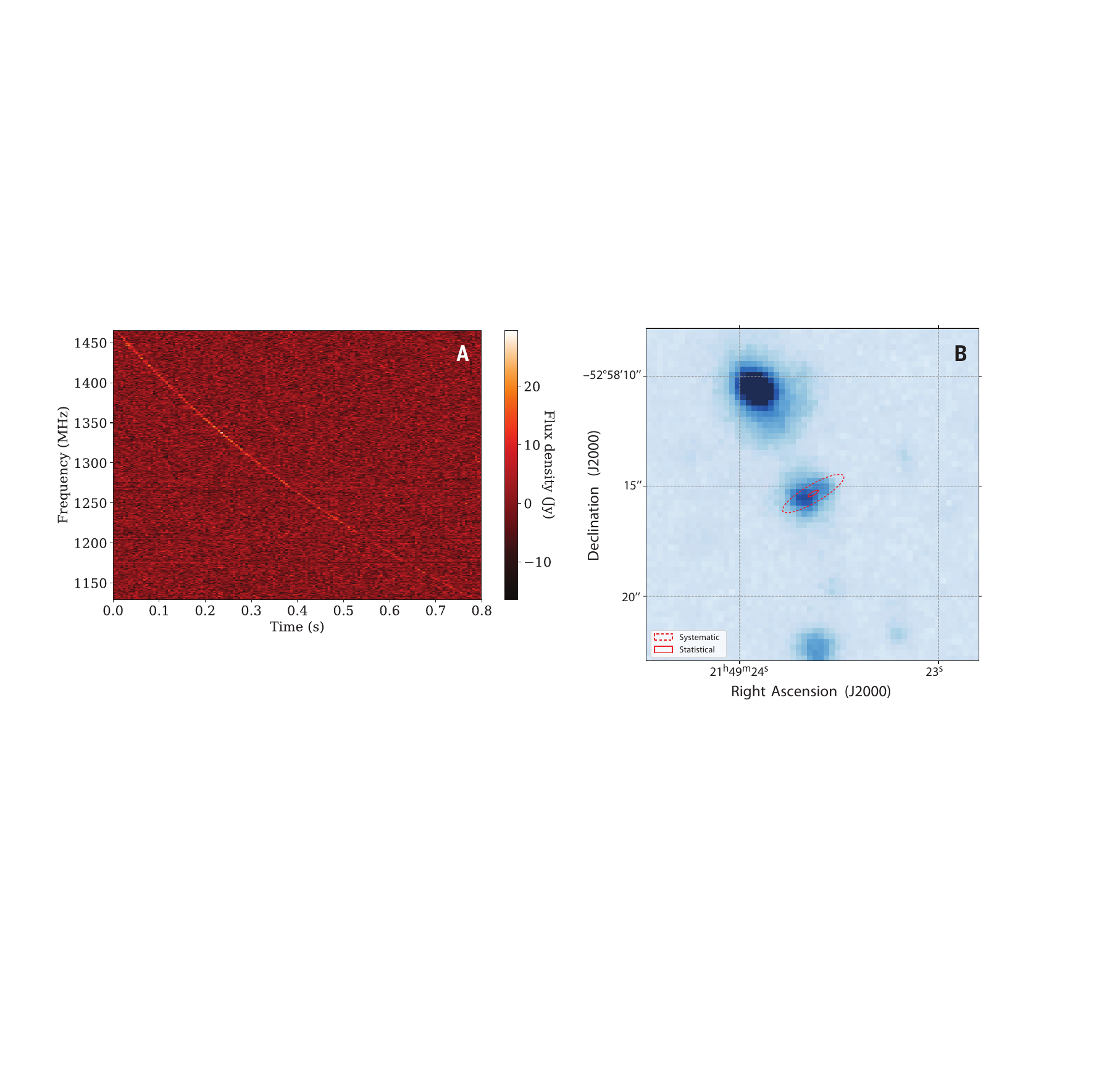}
\vskip -2.5in
\end{center}
    Figure \figfrb: {\bf Dynamic spectrum of FRB~181112
    and optical imaging of its host and a coincident
    foreground galaxy.}
    (A) Dynamic spectrum of FRB~181112 recorded by ASKAP.
    The dispersion measure $\mdmfrb = \dmval \; \mdmunits$;
    (B) $g$-band FORS2 image centered on FRB~181112 whose position
    is depicted by the red ellipses with
    solid/dashed lines indicating the statistical/systematic
    uncertainty.  We estimate an additional systematic
    uncertainty of $\approx 0.5''$ in the astrometric
    solution of the FORS2 image.
    The host is well-localized to a faint
    galaxy cataloged as \host,
    and one identifies a brighter galaxy 
    located $\approx 5''$ away at a PA~$\approx 13^\circ$
    (cataloged as \fgdes, referred to as \fgname).
    The sightline to FRB~181112 passes through the halo
    of this foreground galaxy at 
    an impact parameter $\mrperp = \rperpv$\,kpc.
\end{figure}


\begin{figure}[h]
\begin{center}
\vskip -1.5in
\includegraphics[width=1.2\linewidth]{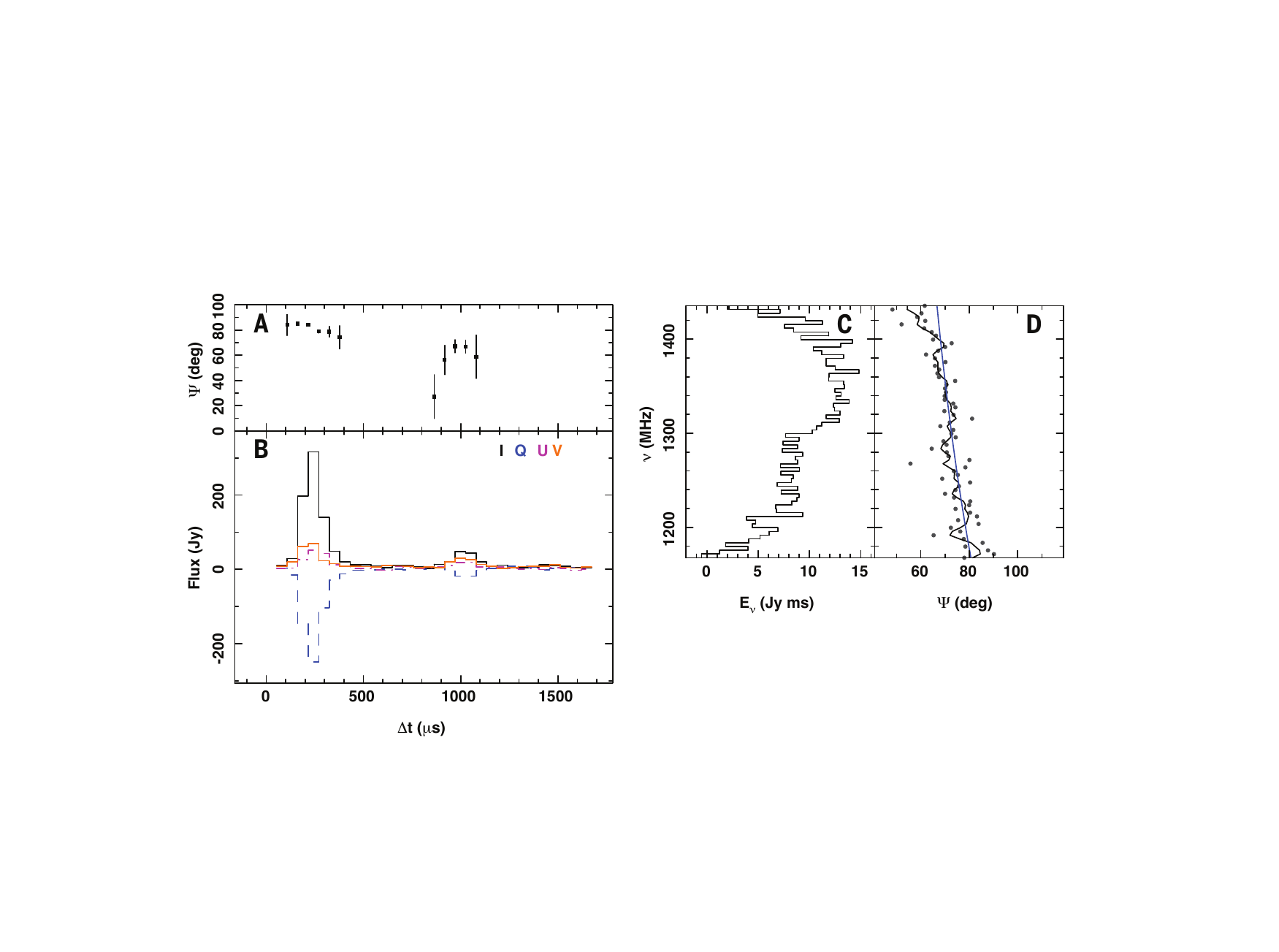}
\vskip -2.5in
\end{center}

    Figure \figrm:  {\bf Spectropolarimeric properties of FRB~181112.}    
    A: Relative linear polarization position angle $\Psi$ of the burst averaged in frequency.  
    B:   Polarimetric pulse profile of burst in four Stokes components (I - thick solid line; Q - dashed line; U - dashed-dotted line, V - thin solid line).   The two components, separated by approximately $800~\upmu$s show different position angles.     C:  Spectrum of $E_\nu$ burst averaged over both pulses.  
    D:  Position angle $\Psi$ of the burst plotted as a function of frequency, with the black points showing measurements in individual frequency channels, and line these measurements smoothed uisng a Gaussian Kernel with standard deviation of 4~MHz.     The variation of the position angle with frequency is the result of Faraday rotation.  The blue line shows a maximum-likelihood model for polarization, using the inferred rotation measure ${\rm RM} = \rmvalue \, \rmunits$. 
\end{figure}

\begin{figure}[h]
\includegraphics[width=\linewidth]{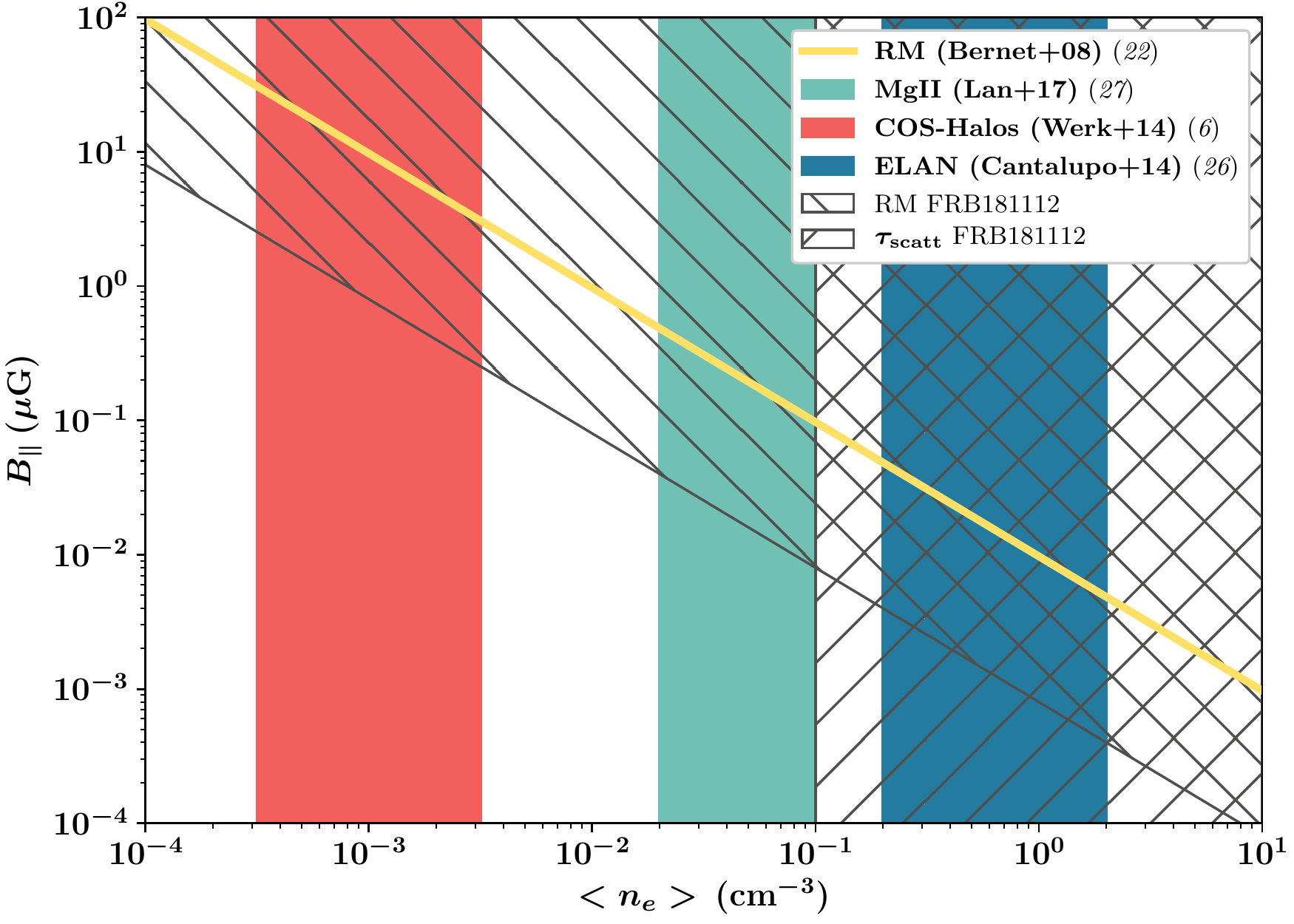}
    Figure \figsumm:
    {\bf Constraints on the coherent magnetic
    field parallel to the line of sight
    $B_\parallel$ and electron density $n_e$
    in the halo of \fgname.} 
    The hatched regions show the parameter space
    in $B_\parallel$, $n_e$ (cool gas) ruled out 
    for the halo of \fgname\ 
    from the measured RM and \dtpulse\ of FRB~181112.
    These constraints are largely
    independent of the properties of the foreground galaxy.
    We compare these results with previous inferences
    for the density of cool halo gas  (colored regions)
    based on ionization modeling and \lya\ flourescence.
    We also illustrate previous estimations for the magnetic
    field strength in halo gas (yellow curve; \cite{bernet08})
    which conflict with our results.
\end{figure}



\begin{table*}
\centering
Table \tabsumm: \caption{{\bf Properties of FRB~181112, its host, and the foreground galaxy \fgname.}
Uncertainties in Right Ascension and Declination
are statistical and systematic, as projected onto
the RA and Decl. axes.  These uncertainties are best described as
ellipses with position angle 120\,deg East of North and major/minor axes
of ($a_{\rm statistical}=0.55'', b_{\rm statistical}=0.15''$) and 
($a_{\rm systematic}=3.2'', b_{\rm systematic}=0.8''$). 
The coherent magnetic field, density and filling factor estimates
assume a characteristic path length through the halo 
of $\Delta L = 50$ kpc.
The density and filling factor estimates assume
a driving scale with root mean density fluctuations of one at $L_0 = 1~$kpc, with the bound scaling as $\propto L_0^{1/3}$, as well as a Kolmogorov spectrum of turbulence to separations below $r_{\rm diff}$.
The filling factor estimate further
assumes cool $T_{\rm cool} = 10^4~$K gas is in pressure equilibrium with hot gas with density $\langle n_e \rangle =10^{-3}$cm$^{-3}$ and temperature $T_{\rm hot}=2\times10^6$K hot gas, with the bound scaling as $\propto (\langle n_e \rangle T_{\rm hot}/ T_{\rm cool})^{-2}$.
See \cite{supp} and the main text for further details. 
}
\begin{tabular}{|l r|}
\hline 
FRB & \\ 
\hline 
Right Ascension (J2000) & $327.34846 \pm 0.00007 \pm 0.0006$ \, deg \\ 
Declination (J2000) & $-52.97093 \pm 0.00004 \pm 0.0002$ \, deg \\ 
Dispersion measure (\dmfrb) & $589.27 \pm 0.03$\, pc cm$^{-3}$ \\ 
Rotation measure (RM) & $10.9 \pm 0.9$\, rad m$^{-2}$ \\ 
Pulse width & $< 40 \, \upmu$s \\ 
\hline 
Host Galaxy & \\ 
\hline 
Redshift & $\zfrb \pm 0.00015$ \\ 
Stellar Mass & $2.6 \pm 1.1$ $\, \times 10^{9} M_\odot$  \\ 
Star formation rate & $0.6$\, \msun \, yr$^{-1}$\\ 
\hline 
Foreground galaxy \fgname & \\ 
\hline 
Redshift & $0.36738 \pm 0.00007$ \\ 
Impact parameter to the FRB sightline ($\mrperp$) & $\rperpv \pm 3$ \, kpc \\ 
Stellar Mass & $4.9 \pm 3.2\, \times 10^{10} M_\odot$ \\ 
Star formation rate & $<0.3\, M_\odot \, \rm yr^{-1}$\\ 
Coherent magnetic field parallel to the line of sight & $B_\parallel <0.5 {\rm \upmu G} \, (n_e/10^{-3} \rm cm^{-3})$ \\ 
Density constraint for hot, diffuse gas ($f_V \sim 1$) & $n_e < 2 \times 10^{-3} \rm cm^{-3}$ \\ 
Filling factor constraint for cool, clumpy gas & $f_V < 10^{-4}$ \\ 
\hline 
\end{tabular} \\ 
\end{table*}


\clearpage

\renewcommand\thesection{S\arabic{section}} 
\setcounter{section}{0} 

\renewcommand\thetable{S\arabic{table}} 
\setcounter{table}{0} 

\renewcommand\thesection{S\arabic{section}} 
\setcounter{section}{0}

\renewcommand\theequation{S\arabic{equation}}
\setcounter{equation}{0}


\begin{center}
{\LARGE
Supplementary Materials for\\}
{\large
 The low density and magnetization of a massive galaxy halo exposed by a fast radio burst
 }
\end{center}
\pagenumbering{gobble}

\noindent 
J. Xavier Prochaska,
Jean-Pierre	Macquart,
Matthew 	McQuinn,
Sunil	Simha,
Ryan  M. Shannon,
Cherie K.	Day,
Lachlan	Marnoch,
Stuart	Ryder,
Adam 	Deller,
Keith W. Bannister,
Shivani	Bhandari,
Rongmon	Bordoloi,
John 	Bunton,
Hyerin	Cho,
Chris	Flynn,
Elizabeth K. Mahony,
Chris	Phillips,
Hao	Qiu,
Nicolas	Tejos

\noindent
{\bf This PDF file includes}\\
Materials and Methods\\
Supplementary Text\\
Figures S1 to S12\\
Tables S1 to S7 \\

\newpage
\pagenumbering{arabic}
    \setcounter{page}{1}

\section{Materials and Methods}

Throughout the paper,
for calculations that require an assumed cosmology
we adopted the model and parameters from the
Planck 2015 analysis \cite{planck15}.

\subsection{Data Acquisition and Processing}
\label{sec:acqpros}

The ASKAP FRB real-time detection system was configured in an identical manner to that described in \cite{Bannisteretal2019}. At the time of detection of FRB~181112, 
twelve of the 36 antennas were operational with a maximum baseline of 4392 m.

The method used to determine FRB~181112's astrometric position and uncertainties follows that described in \cite{Bannisteretal2019}. Briefly, four sets of voltage data were captured from the ASKAP telescope.  The duration of each voltage data capture was 3.1s and the total bandwidth was 336MHz, comprised of $336 \times 1$~MHz complex-sampled dual polarisation sub-bands, with an over-sampling ratio of 32/27 and 4-bit quantisation precision.  The voltage captures occurred at the following times and positions, where the positions are the J2000 RA\ and Decl.\ at the beam centre:
\begin{enumerate}
\item FRB~181112: $\rm RA, Decl. = 21h50.0m57.7s$, $\rm -53d10'32''$, central time UTC~2018-11-12:~17:31:16.6324205+00:00
\item PKS~0407$-$658 (visit~1), a bright, compact calibrator source: $\rm RA, Decl. = 04h08m20.4s, -65d45'09''$, $\sim$5 hours after the FRB [UTC~2018-11-12:~22:30:10.4107085+00:00])
\item PKS~0407$-$658 (visit~2): $\rm RA, Decl. = 04h08m20.4s, -65d45'09''$, central time UTC~2018-11-12:~23:55:04.4360995+00:00)
\item PSR~J0834$-$4510, the Vela pulsar: $\rm RA, Decl. = 08h35m20.7s$, $\rm -45d10'35''$, central time UTC~2018-11-13:~00:43:55.610311+00:00)
\end{enumerate}

The first two voltage datasets were affected by a hardware issue which resulted in the loss of 2.3\% of the data from one antenna. Shortly after this issue was discovered, at approximately UTC~2018-11-12:~23:00, some components of the ASKAP hardware were reset in order to fix this issue in further observations.  This hardware reset meant that calibration solutions from the PKS~0407$-$658 (visit~1) dataset could not be applied to the subsequently observed Vela dataset, and accordingly the PKS~0407$-$658 (visit~2) dataset was also acquired.  
From these four sets of voltage data, nine visibility datasets were produced using the software correlator package {\sc DiFX} \cite{Deller11}. In the summary below, the time resolution refers to the maximum time difference between the first and last data added into an individual visibility accumulator for a given timestep, while the total effective integration time refers to the total amount of valid data added into all visibility accumulators for the entire correlation.  
The nine datasets are as follows:
\begin{itemize}
    \item PKS~0407$-$658 (visit~1) data, correlation centred at PKS~0407$-$658 ($\rm RA, Decl. = 04h08m20.380s$, $\rm -65d45'09.08''$), time resolution 1.3824~s, total effective integration time 3.1~s, correlator frequency resolution 9.3~kHz (the ``FRB calibrator" dataset)
    \item FRB~181112 data, correlation centred at the approximate FRB position ($\rm RA, Decl. = 21h49m20s, -52d58'00''$), time resolution 1~ms, total effective integration time 1~ms, correlator frequency resolution 9.3~kHz (``FRB gated" dataset)
    \item FRB~181112 data, correlation centred on the ASKAP beam centre \break ($\rm RA, Decl. = 21h50m57.69s, -53d10'32.48''$), time resolution 1.3824~s, total effective integration time 3.1~s, correlator frequency resolution 9.3~kHz (``FRB continuum field" dataset)
    \item FRB~181112 data, correlation centred near the final FRB position ($\rm RA, Decl. = 21h49m23.6s$, $\rm -52d58'15.4''$), time resolution $\rm 54\mu s$, total effective integration time 1.62~ms, correlator frequency resolution 18.5~kHz  (``FRB high time resolution" dataset)
    \item FRB~181112 data, correlation centred near on the FRB position ($\rm RA, Decl. = 21h49m20s$, $\rm -52d58'00''$),  time resolution 21~ms, total effective integration time 16~ms, correlator frequency resolution 9.3~kHz (``FRB radio-frequency interference (RFI) subtraction" dataset)
    \item FRB~181112 data, correlation centred near on the FRB position ($\rm RA, Decl. = 21h49m23.6s$, $\rm -52d58'15.4''$),  time resolution 21~ms, total effective integration time 16~ms, correlator frequency resolution 18.5~kHz (``FRB high time resolution RFI subtraction" dataset)
    \item PKS~0407$-$658 (visit~2) data, correlation centred at PKS~0407-658 ($\rm RA, Decl. = 04h08m20.380s$, $\rm -65d45'09.08''$), time resolution 1.3824~s, total effective integration time 3.1~s, correlator frequency resolution 9.3~kHz  (``Vela calibrator" dataset)
    \item Vela data, correlation centred at the position of Vela ($\rm RA, Decl. = 08h35m20.65525s$, $\rm -45d10'35.1545''$) time resolution 1.3824~s, total effective integration time 62.3~ms, correlator frequency resolution 9.3~kHz (``Vela" dataset)
    \item Vela data, correlation centred at the position of Vela ($\rm RA, Decl. = 08h35m20.65525s$, $\rm -45d10'35.1545''$) time resolution 1.3824~s, total effective integration time 311.5~ms, correlator frequency resolution 9.3~kHz (``Vela RFI subtraction" dataset)
\end{itemize}

\noindent
All datasets were averaged after correlation to a frequency resolution of 167 kHz.

PKS~0407$-$658 was first observed $\sim5$ hours after FRB~181112 (vist~1) and again 
$\sim6.4$\,hours later (visit~2). The RFI free part of the band ($\sim$ 266\,MHz and 238\,MHz, respectively, concentrated in the upper half of the band) was used to determine antenna-based delays, which were then applied to both the calibrator correlation and the associated target field (FRB~181112, or Vela) correlation. Calibration solutions for all datasets were derived using the \textit{AIPS} \cite{Greisen03} tasks FRING and CALIB on their respective calibrator datasets. The instrumental bandpass was corrected using the \textit{AIPS} task CPASS, which interpolates over the RFI flagged region of the band when deriving these frequency-dependent phase and amplitude corrections.

As described in \cite{Bannisteretal2019}, we produced a second set of visibilities for each of the FRB gated, FRB high time resolution, and Vela datasets, to remove the effects of RFI. These datasets are referred to above as ``RFI subtraction'' datasets. These visibilities were formed by correlating and integrating over two windows on either side of the ``on-pulse'' window in time, with a gap between the ``on'' and ``off'' pulse regions. In the pulsar binning mode used to produce the FRB and Vela datasets (and their RFI subtraction datasets), the DiFX correlator scales the visibility amplitudes by the bin width.  Accordingly, to normalise the two data sets to the same amplitude scale, the RFI subtraction visibilities were scaled by the ratio $t_{\rm on} / t_{\rm off}$, where $t_{\rm on}$ and $t_{\rm off}$ are the effective integration times of the on- and off-pulse datasets, respectively. These scaled datasets were then subtracted from their ``on-pulse'' counterparts to remove any contributions to correlated ``on-pulse'' visibilities that are approximately constant on timescales of 20~ms (which include both RFI and any background celestial sources). This scaling and subtraction was accomplished via a custom ParselTongue \cite{Kettenis06} script 
({\sc uvsubScaled.py}). 

\subsection{Radio Imaging and Astrometric Position Extraction of FRB~181112 and the field}
\label{sec:astrom}

To determine the FRB position, we imaged  the gated FRB dataset in \textit{CASA}  \cite{CASA}, using the task TCLEAN in widefield, multifrequency synthesis, multiscale mode with natural weighting, producing a 256 $\times$ 256 pixel image with 2.5 arcsec pixels. The FRB~181112 position and its uncertainties were then extracted using the \textit{AIPS} task JMFIT on the Stokes I image, producing a detection with a significance of 37$\sigma$.

The custom {\sc Miriad} \cite{miriad} script described in \cite{Bannisteretal2019} was then used to similarly reweight the FRB gated dataset's visibilities, normalizing the FRB amplitude across the band by up- and down-weighting the brightest and faintest regions, respectively, and adjusting the visibility weights by the inverse of the amplitude correction. This reweighted dataset was imaged with TCLEAN, using the same settings as the unweighted dataset, and its position and uncertainties determined with JMFIT to be $\rm RA, Decl. = 21h49m23.63s \pm 0.05s, -52d58'15.4'' \pm 0.3''$, with a detection significance of 54$\sigma$ and offsets in RA and Decl. from the previous, naturally-weighted dataset that are less than the positional uncertainties measured for either weighting scheme.

To determine systematic uncertainty in the ASKAP astrometric positions, the FRB continuum field dataset was flagged to remove visibilities corrupted by RFI or dish shadowing (the FRB was observed at low elevation ($14.2$~deg), producing shadowing on some very short baselines) and then imaged in \textit{CASA}.  The same TCLEAN settings were used as for the FRB gated dataset, except that a 3000 $\times$ 3000 pixel image with 2.5 arcsec pixels was produced. Six background radio sources were detected, with only one having a significance above 10$\sigma$. For each of the detected background sources, we used JMFIT to determine its position and uncertainty in the ASKAP image.

To tie the ASKAP field source positions to the third International Celestial Reference Frame (ICRF3) and to search for any radio continuum emission associated with the FRB host galaxy, a 12-hour observation of the FRB field was carried out with the Australia Telescope Compact Array (ATCA) in the 6B array configuration on 2018~Nov~17 UTC.  This included scans of three of the background sources described above (SUMSS~215155$-$531646, \break SUMSS~215606$-$532916, and SUMSS~215036$-$525538, see Table~\ref{tab:ASKAPsrcs}) at 1.1 - 3.1 GHz (with heavy RFI flagging from 1.1 - 1.5 GHz) interleaved with observations of three ATCA calibrators (with only two being used for the final analysis, see Table~\ref{tab:ATCAcal}), as well as 4.5 - 8.5 GHz observations centered on the FRB position. 

Using multi-frequency synthesis, we combined all of the $4.5 - 8.5$ GHz data on the FRB field, producing an image with no detection at the FRB position and a 3$\sigma$ upper limit of 21~$\upmu$Jy.  

The $1.1 - 3.1$~GHz observations were used to refine the positions of the  three ASKAP background sources. However, SUMSS~215036$-$525538 was found to be too faint to contribute a substantial constraint on the absolute ASKAP accuracy, while SUMSS~215606$-$532916 was found to be resolved in the ATCA images, also precluding an accurate positional comparison. Consequently, only SUMSS~215155$-$531646 was used to estimate any systematic errors in the ASKAP image. The position for SUMSS~215155$-$531646 has a statistical precision of $\sim$10~milliarcseconds, but the position differs by $\sim$0.5 arcsec when obtained using PKS~B2117$-$642 versus PKS~B2215$-$508 as the calibrator source (see Table~\ref{tab:ATCAcal}.) We use the average of the two positions so obtained as the final estimate of the ATCA position for SUMSS~215155$-$531646 and take the difference between them as an estimate of the uncertainty on this position. While not negligible, this discrepancy is still small compared to the statistical precision to which the ASKAP position for SUMSS~215155$-$531646 is measured. 
The final offset between the ATCA and ASKAP positions for SUMSS~215155$-$531646 was found to be $-0.61 \pm 2.15$ arcsec in RA and $0.68 \pm 1.35$ arcsec in Decl. Accordingly, as was the case for FRB~180924 \cite{Bannisteretal2019}, our astrometric checks are consistent with no systematic offsets in the ASKAP astrometry, but the precision to which this is confirmed is much lower than for FRB~180924 due to the fainter background sources and the reduced number of antennas active when the FRB was detected. Combining the FRB measurement error with the estimated systematic uncertainty yields a final FRB position of $\rm RA, Decl. = 21h49m23.63s \pm 0.24s, -52d58'15.4'' \pm 1.4''$. 
Due to the elongated beam shape, the uncertainty in RA and Decl. is highly correlated for both the statistical and the (dominant) systematic contributions.  We report the uncertainty projected onto the RA and Decl. axes as is customary, 
but Fig.~\figfrb\ shows the total uncertainty ellipse, as generated by combining 
in quadrature the statistical and systematic positional uncertainty ellipses from JMFIT.  As the source being fitted is unresolved in both cases, the size and position angle of the Gaussian model is approximately that of the ASKAP synthesised beam (16.5$\times$60.0$''$ at p.a. 120 degrees East of North), with the uncertainty ellipse being a scaled version 
of the model.




\subsection{Spectropolarimetry:  calibration and rotation measures}

To study the polarimetric properties of the burst (both the spectral variations associated with Faraday rotation and temporal variations) the burst data products needed to be calibrated. 
The calibration strategy used follows methods developed for the 
localized ASKAP burst described in \cite{Bannisteretal2019}, which we summarize here.

The largest calibration defect is leakage between Stokes U and V due to an unknown phase $\Phi_0$ and delay $\Delta \tau$ between the two orthogonal linear polarization receptors.
We do not attempt to correct for the absolute position angle of the burst on the sky. 

We can measure these terms by observing a source with determined polarization properties.  As in previous work \cite{Bannisteretal2019}, we used observations of the Vela pulsar (PSR~J0835$-$4510), assuming that it has the following fractional polarization properties \cite{Bannisteretal2019}:
\begin{eqnarray}
&&\frac{L}{I}(\nu) = \frac{\sqrt{Q^2(\nu) + U^2(\nu)}}{I(\nu)} = 0.95\\
&&\frac{V}{I}(\nu) = -0.05 \\
&&\frac{Q}{I}(\nu) = \frac{L}{I}(\nu) \cos\left( 2 \rm{PA}(\nu)\right) \\
&&\frac{U}{I}(\nu) = \frac{L}{I}(\nu) \sin \left( 2 \rm{PA}(\nu)\right),\\
\end{eqnarray}
and the  polarization position angle is
\begin{equation}
\rm{PA}(f) =  0.09\,{\rm RM}(\nu_{\rm GHz}^{-2} - 1.4^{-2}) - PA_0.
\end{equation}
Here ${\rm PA}_0$ is the polarization position angle at a reference frequency of 1.4 GHz.  
Based on Parkes observations reported previously \cite{Bannisteretal2019}, we find RM = $40 \pm 1$ rad\,m$^{-2}$  and ${\rm PA}_0=-0.35$ rad.   

We can then solve for the leakage terms ($\Phi_0$,$\Delta \tau$ and $\Psi$) from the uncorrected fractional linear polarization:  
\begin{eqnarray}
&&\frac{U^{\prime}}{I}(\nu)   = \frac{L}{I}(\nu) \sin\left( 2 {\rm PA}(\nu) + \Psi_0 + 2\pi \nu \Delta \tau + \Phi_0 \right) + \frac{V}{I}(\nu) \sin\left(2 \pi \nu \Delta \tau + \Phi_0\right)\\
&&\frac{V^\prime}{I}(\nu) = -\frac{L}{I}(\nu) \sin \left(2 \rm{PA(\nu)} + \Psi_0 + 2 \pi \nu \Delta \tau+ \Phi_0  \right) + \frac{V}{I}(\nu) \cos\left(2 \pi \nu \Delta \tau + \Phi_0 \right) 
\end{eqnarray}

After correcting $U$ and $Q$ for the leakage terms, uncertainties on the relative  polarization position angle were determined using standard techniques \cite{EverettWeisberg01}.

\subsubsection{Calculating the rotation measure}

Bayesian methodology is used to calculate the rotation measure for the FRB, using the data as calibrated above and a technique applied previously \cite{Bannisteretal2019}.
To summarize, we measured the rotation measure by modelling the fractional polarization in the calibrated $Q_i$ and  $U_i$ for each channel $i$.

We model the Stokes parameters $\hat{Q}_{m,i}$ and $\hat{U}_{m,i}$ in each channel $i$ to be
\begin{equation}
\hat{Q}_{m,i}=L_i \cos\left(2 {\rm PA}_{\rm FRB}(\nu_i)\right)
\end{equation}
and
\begin{equation}
\hat{U}_{m,i}=L_i \sin\left(2 {\rm PA}_{\rm FRB}(\nu_i)\right),
\end{equation}
where the position angle is 
\begin{equation}
{\rm PA}_{\rm FRB}(\nu_i) = 0.09\,{\rm RM}_{\rm FRB} \nu^{-2}_{i, GHz} + {\rm PA}_0,
\end{equation} 
and $L_i$ is the linearly polarized flux density.
We assume that the noise is identical in $Q$ and $U$, is equal across the band and has a Gaussian distribution.

 The likelihood in this case is:
\begin{equation}
\mathcal{L} = \prod_{i}^{N_c} \frac{1}{2 \pi \sigma^2} \exp \left[- \frac{(Q_i-\hat{Q}_{m_i})^2}{2 \sigma^2} \right]    \exp \left[-\frac{(U_i-\hat{U}_{m_i})^2}{2 \sigma^2} \right] 
\end{equation}

We calculate the posterior probabilities for the parameters (${\rm PA}_0$, ${\rm RM}_{\rm FRB}$, and $\sigma$).  
The total polarized intensities in all channels $L_i$ are nuisance parameters over which we can analytically marginalize, assuming uniform priors.  
The final likelihood is then
\begin{equation}
\mathcal{L} = \prod_i^{N_c}  \frac{1}{2 \pi \sigma^2} \exp \left[- \frac{Q_i^2 + U_i^2 -\left( Q_i \cos(2{\rm PA}_{\rm FRB}) + U_i \sin (2 {\rm PA}_{\rm FRB})  \right)^2  }{2 \sigma^2} \right].
\end{equation}

We assume uniform priors on ${\rm PA}_0$, ${\rm RM}_{\rm FRB}$, and  Gaussian priors on  $\tau$  and $\Phi_0$, with the means of $\tau$ and  $\Phi_0$  set to their maximum-likelihood values from the model fitted to the Vela pulsar 
and the standard deviations set to the best-fitting uncertainties. We sample the posterior distribution using a nested-sampling algorithm \cite{Feroz09}.   
We recover RM$_{\rm FRB} = \rmvalue \pm \rmsig \, \rmunits$.

\subsection{Optical and near-infrared Observations}
\label{sec:obs}

\noindent
{\bf Photometry:}
We have collated cataloged photometry from the
Dark Energy Survey (DES) Data Release 1 (DR1; \cite{desdr1})
and  the Wide-field Infrared Survey Explorer (WISE; \cite{Wright10})
all-sky survey of the host galaxy of FRB~181112 
and \fgname\ for
spectral energy distribution (SED) analyses.  
These data are listed in Table~\stabphotom.

We also obtained deeper $g$ and $I$-band images
of the field with the FORS2 imaging spectrograph \cite{FORS2} on 
the ESO VLT in Service Mode on 2018 Dec 3 UT. Five exposures of
500~s in the g\_HIGH filter with incremental offsets of
10$^{\prime\prime}$ between each were followed by 5$\times$90~s
exposures in the I\_BESS filter with similar offsets. The Standard
Resolution collimator was used with 2$\times$2 binning of the CCD
array to yield a pixel scale of 0.25$^{\prime\prime}$ per pixel.
Skies were photometric, and the median seeing was
FWHM~$\approx 0.6^{\prime\prime}$.

Standard image processing was performed using version 2.9.1 of the
ESO Reflex pipeline. The individual images in each filter were 
combined with the {\sc montage} software
package \cite{berriman17}. Astrometric
calibration was applied using
Astrometry.net \cite{lang10} with
a precision of $\approx 0.3''$ when compared to the DES catalog, while
photometric calibration was derived from DES DR1 for the
$g$ data, or from the ESO night monitoring 
data
for the $I$-band.
Integrated magnitudes for the host galaxy of FRB~181112 and \fgname\ were then extracted from the images using the \texttt{flux\_auto} quantity provided by SExtractor 2.19.5 \cite{sextractor}.
For the host galaxy, we find $g = 22.57 \pm 0.04$ and $I = 21.51 \pm 0.04 $; for \fgname, $g = 21.20 \pm 0.02$ and $I = 19.20 \pm 0.02$.
Fig.~{\figfrb}B shows a part of the $g$-band image.

\noindent
{\bf Spectroscopy:}
On 2018 Dec~5 UT we obtained a series of three 890~s 
spectral images of \fgname\ and the host of FRB~181112
with the VLT/FORS2 spectrograph, configured with the 
GRIS\_300I grism and a $1''$ wide slit 
(full-width half max resolving power, $R_{\rm FWHM} \approx 660$) 
oriented at a position angle
of 24$^{\circ}$ to include both galaxies simultaneously.
These and associated calibration images were processed 
with the PypeIt software package \cite{pypeit} 
and the spectra were optimally extracted and flux-calibrated
with standard procedures.  
The latter includes a correction to match the flux integrated
through the DES $r$ filter to its cataloged magnitude
(Table~\stabphotom).
The data were taken at an airmass of $\approx 2$ with the slit
well away from the parallactic angle, and
we have not corrected for losses from differential refraction.

\subsection{Properties of the Host Galaxy of FRB~181112}
\label{sec:host}

Although not the primary focus of this manuscript, we report
here the properties 
of the host galaxy as these may constrain 
progenitor models of FRBs (e.g.\ \cite{Plattsetal2018}).
For the radio data, the non-detection of emission in the 4.5 - 8.5 GHz ATCA image corresponds to an upper limit of $L_{(6.5 GHz)} < 1.3 \times 10^{22}$ W/Hz.  Accordingly, if the host of FRB 181112 were to harbor a persistent radio source of the kind seen co-located with the repeating FRB 121102 (luminosity $\sim$ $1.8\times10^{22}$ W/Hz), it must be fainter than the one associated with FRB 121102.

For the optical data, Fig.~\sfighostcigale\ shows the best fit SED to the 
DES photometry of the host galaxy (Table~\stabphotom) performed
with the CIGALE software package \cite{Noll:2009aa}.
The CIGALE package requires an input file that defines ``modules'' 
for fitting along with initial guesses for module parameters. 
Each parameter is allowed to take values in the provided initial guess array. 
CIGALE then computes SEDs for all possible combinations of input 
parameters in the multidimensional grid defined by the input guesses. After computing the 
$\chi^2$ for all models, CIGALE performs a Bayesian analysis and outputs the 
fit parameters that correspond to the model with the highest likelihood. 
In the case of the host galaxy, we made use of the following modules and constraints:
\begin{enumerate}
    \item A delayed-exponential SFH model with no late burst population.
    \item The \cite{Bruzual:2003aa} simple stellar population with Chabrier initial mass function and metallicity allowed to vary from 0.005$Z_\odot$ to 2.55$Z_\odot$.
    \item The \cite{Calzetti:2000lr} dust extinction 
    model with extinction computed 
    for the DES $g$ and $r$ bands.
    \item The \cite{Dale:2014aa} dust emission model with an AGN fraction 
    $f_{\rm AGN} \le 0.05$.
\end{enumerate}
The galaxy is not detected in existing near or far-infrared datasets
(e.g.\ WISE) and therefore the dust emission is poorly 
constrained. CIGALE's results are summarised in Table~\stabhost\, 
including the estimated stellar mass
$M_{*, \rm host} = 10^{\hostmstar \pm \hostsigmstar} \mmsun$,
star-formation rate
SFR~$= \hostsfr \pm \hostsigsfr \mmsun \, \rm yr^{-1}$, 
and internal reddening
$E(B-V) = \hostebv \pm \hostsigebv$\,mag. The large uncertainties
arise from the 
degeneracies that are inherent to SED modeling (e.g.\ constraining a 
large number of fitting parameters with relatively few data points). 

Our FORS2 spectrum also permits a spectral analysis to assess
the galaxy's stellar population and its associated nebular emission.
We analyzed these data with the pPXF software package \cite{Cappellari:2017aa}
and a portion of the best-fit model is shown in Fig.~\sfighostppxf.
The signal-to-noise
of these data are too low to provide precise constraints on
the stellar population,
nevertheless the estimated mass-to-light ratio is
consistent with that derived from the photometry.
From the observed line fluxes (Table~\stabflux), we assess 
the nature of the nebular line-emission from the galaxy.
Unfortunately, the H$\alpha$ and [N{\sc ii}] lines
lie at $\approx 9700$\AA\ where telluric absorption is
substantial, especially at the high airmass of the observations.
We have therefore inflated the uncertainty estimates
to $3 \times 10^{-17} \rm \, erg \, s^{-1} \, cm^{-2} \, Ang^{-1}$.
Placing the [O{\sc iii}]/H$\beta$ and [N{\sc ii}]/H$\alpha$ line-ratios
on the  Baldwin-Phillips-Terlevich 
(BPT) diagram yields an inconclusive result,
although the more precise [O{\sc iii}]/H$\beta$ ratio implies 
a lower metallicity galaxy which is consistent with its low
stellar mass.
The large systematic uncertainty in the H$\alpha$ flux also precludes
a precise estimate of the internal extinction from the
measured Balmer line ratios.  
Instead, we adopt the $E(B-V) = 0.15$\,mag
derived from the photometric analysis (Table~\stabhost).
From the H$\beta$ luminosity (Table~\ref{tab:lines}), 
we estimate the star-formation rate,
SFR$_{\beta} = \hostsfrbeta \mmsun \, \rm yr^{-1}$,
consistent with the estimate from photometry. 

Fig.~\sfighosts\ places the host galaxy
in context with the other
three well-studied FRB host galaxies \cite{Tendulkar17,Bannisteretal2019,Ravietal2019},
and the general population
of $z \sim 0.3$ galaxies \cite{primus}.
We find that the host galaxy has a mass intermediate between 
the hosts of FRB~1211102 and FRB~180924
and that it lies on the locus of 
star-forming galaxies.
Therefore, our observations
disfavor models that invoke dwarf galaxies
as the dominant host of FRBs, and also
scenarios dominated by 
super-luminous supernovae usually found in low luminosity/metallicity galaxies \cite{nicholl+17}.

\subsection{Associating FRB~181112 with \host}
\label{sec:association}
While FRB 181112 appears to be associated with \host~ based on the localization, 
it is possible that this is simply a chance alignment on the sky.
To explore this possibility, we have
estimated the probability of a chance alignment as follows.

We take a set of \npatch\ random 1 sq. deg. patches 
of sky within the 
DES DR1 field and cover it with a uniform square grid 
with $1''$ spacing. 
We then compute the fraction of the grid with centers that fall within $1''$ of a cataloged DES
galaxy with $r<22$\,mag. 
This fraction has an average value
of 0.31\% (see Fig.~\sfigrandom). 
The chance of randomly finding two such FRB-host pairs with ASKAP
(FRB181112, presented here, and FRB180924, from \cite{Bannisteretal2019}) is 
therefore on the order of $10^{-5}$.
We conclude that FRB~181112 is associated with
the luminous galaxy \host.

\subsection{Properties of \fgname}
\label{sec:fg_galaxy}

Similar to the host galaxy, we have performed an SED analysis of 
\fgname\ from the collated photometry (Table~\stabphotom).
We used the same modules as in the case of the host but allowed 
a maximum AGN fraction of 0.1.  The best-fit model
is shown in Fig.~\sfigfgcigale\ and summarized in Table~\tabsumm.
The results indicate a massive galaxy
($\log_{10} \mmstar/\mmsun = 
\logmstar^{+\usiglogm}_{-\lsiglogm}$),  with
internal reddening
$E(B-V) = \fgcebv$\,mag, and a high metallicity
$Z_{\rm fg}= \fgcmetal$ with ${\rm Z}_\odot$ the solar
metallicity. 

The optical spectrum (Fig.~\sfigfgspec)
was further analyzed with the pPXF software
package \cite{Cappellari:2017aa} to perform line-flux measurements
(including the effects of Balmer absorption) and to assess
the underlying stellar population.
This analysis, summarized in Table~\tabsumm, yields an estimate
of $E(B-V) = \fgebv$ and a metallicity $Z_{\rm fg} = \fgmetal$
(relative to Solar). 

The line flux ratios of [O{\sc iii}]/H$\beta$ and
[N{\sc ii}]/H$\alpha$ place this galaxy in the Seyfert 
regime of the BPT diagram (Fig.~\sfigbpt), 
and we place an upper limit
on the SFR based on the 
Balmer-corrected H$\alpha$ flux of 
SFR$_\alpha < \fgsfr \, M_\odot \, \rm yr^{-1}$ (see also Table~\tabsumm). 
We used this upper bound to constrain our CIGALE analysis. 


\subsection{Morphology of the Host and the Foreground galaxies}

S\'{e}rsic profile fits in the $I$-band image 
were made of the host and foreground galaxies simultaneously 
because of their proximity in the sky. The fits were made using 
GALFIT \cite{galfitv3} 
assuming three components:
(i) a constant sky background,
(ii) a S\'{e}rsic profile for the foreground galaxy and
(iii) a S\'{e}rsic profile for the host galaxy.
The point-spread function (PSF) was estimated from 
unsaturated sources consistent with circular images using the \texttt{photutils} package \cite{photutils}.

The results of the fitting
are shown in Fig.~\sfigsersic\ and listed in Table~\stabsersic. 
Inputs to GALFIT include an estimate of the PSF, the image, 
and a list of components (with initial guesses for parameters) to be fitted. GALFIT obtains the best fitting parameters in the minimum $\chi^2$ sense. 
The parameters are sensitive to the fitting area, especially the S\'{e}rsic indices of the two galaxies. In the case of an isolated galaxy, as the area increases, GALFIT tends to assign higher indices i.e. steeper profiles until a critical area beyond which it plateaus. This steady value is accepted as the light profile's true index.

There is, however, a certain degree of systematic uncertainty involved. 
One cannot simply increase the fitting area indefinitely, and usually one encounters other objects and must terminate the fitting process. 
We varied the fitting area to see how the parameter values are affected and we 
are confident that the fitted uncertainties quoted by GALFIT adequately bound the values.
The PA derived indicates the sightline to FRB~181112
occurs approximately along the major axis of \fgname.

\section{Supplementary Text}

\subsection{The DM of FRB181112}
\label{sec:DM}

The intersection of an FRB sightline with a foreground, massive
galactic halo is expected to be a rare occurrence.
Applying standard techniques \cite{moster10} based 
on the redshift and stellar mass, 
we estimate the galaxy's
dark matter halo mass to be $\mmhalo \sim 10^{\mhalov} \, \mmsun$.
Adopting the mass function 
from the Aemulus simulations \cite{AEMULUSI},
the comoving number density of such halos with 
$M \ge \mmhalo$ at $z=0.367$ is 
$n(M \ge \mmhalo; z=0.367) \approx 10^{-3} \, \rm Mpc^{-3}$.
The probability to intersect any such halo per absorption length
$dX$ within an impact parameter $\mrperp \le \rperpv$\,kpc is

\begin{equation}
\ell(X) = \frac{c}{H_0} n \, \pi \mrperp^2
\end{equation}
with $H_0$ Hubble's constant,
$c$ the speed of light,
$n$ the number density of halos with $M > \mmhalo$
in comoving units and $\mrperp$ in physical units, and

\begin{equation}
dX \equiv \frac{H_0}{H(z)} \, (1+z)^2 \, dz \;\;\; .
\end{equation}
Integrating $\ell(X) dX$ from $z=0$ to \zfrb, we 
obtain the mean number of halos within the volume of interest (assuming these are randomly distributed). Then, from Poisson statistics, we estimate the chance of intersecting one halo to be $\approx 0.5 \%$.

Given the rarity of this event and the high mass of
\fgname, one might expect its halo gas to dominate the
dispersion measure
DM$_{\rm FRB}$ of the FRB.  To explore this hypothesis, we have
constructed a DM probability distribution function (PDF)
for a set of 10,000 FRBs at $\mzfrb = \zfrb$
following the formalism of \cite{xyz19}.
Specifically, we have assumed the Galactic interstellar
medium (ISM) contributes
DM$_{\rm ISM} = 42\, \mdmunits$ \cite{NE2001},
the Galactic halo contributes
DM$_{\rm MW,halo} = 60\, \mdmunits$ \cite{xyz19},
and the host contributes
DM$_{\rm host} = 32\, \mdmunits$ (this assumes an intrinsic
host ISM + halo DM of 50 before evaluating it at $z_{\rm FRB}=\zfrb$).
For the cosmic contribution (the intergalactic medium
IGM plus intervening halos) to DM$_{\rm FRB}$, we have
assumed that galactic halos with $\mmhalo \ge 10^{11} \mmsun$
have retained $f_b = 0.75$ of the
cosmic fraction of baryons (and 0\%\ otherwise)
as an ionized halo gas, and that
these are distributed with a modified Navarro-Frenk-White
profile \cite{nfw97}
(mNFW, with $\alpha=2, y_0=2$; see \cite{xyz19}).  This means
that approximately half of the baryons today are 
locked in galactic halos and the remaining fraction $f_{\rm IGM}$ is
located in the more diffuse IGM.
We may express the DM for the diffuse IGM as:

\begin{equation}
{\rm DM}_{\rm IGM} =  \int ds \, f_{\rm IGM} \, \bar n_e / (1+z)  
\end{equation}
where  $\bar n_e = \rho_b(z) \mu_e / \mu_m m_p$ and where
$f_{\rm IGM} (z)$ is the fraction of cosmic baryons in the diffuse IGM,
$\rho_b \equiv \Omega_b \rho_c$,  and $\mu_m = 1.3$ 
and $\mu_e=1.1667$ account for the mass and electrons 
of Helium \cite{xyz19}.
We then generated random realizations of halos foreground to 
FRB~181112 and created the PDF shown in Fig.~\sfigDMpdf.
It is evident that the DM of FRB~181112 -- corrected for the 
Galactic ISM, Galactic halo, and host galaxy -- lies
below the median value of the PDF, i.e.\ it is not an
anomalously high DM event.

This conclusion is further illustrated in Fig.~\sfigDMcum\ which
shows the integrated DM for FRB~181112 from Earth to the event.
For the cosmic DM, we have only included DM$_{\rm IGM}$ and our
estimate for the DM of \fgname's halo.
The latter assumes $\mmhalo = 10^{\mhalov} \mmsun$,
$\mrperp = \rperpv$\,kpc, $f_b = 0.75$, and the
modified NFW profile yielding DM$_{\rm FG} = \dmfg \, \mdmunits$
after down-weighting by $(1+z_{\rm fg})^{-1}$.
This analysis suggest that there need not be any
additional massive halos in the
foreground of FRB~181112 that contribute to DM$_{\rm FRB}$.
The halo gas model adopted in the
analysis (mNFW) assumes a higher baryon fraction ($f_b$)
retained in galactic halos than results from some numerical
simulations (e.g.\ \cite{fielding+17}).
Adopting the \cite{fielding+17} density profile scaled
to a halo with $\mmhalo = 10^{\mhalov} \mmsun$
gives DM$_{\rm FG} \approx 50\, \mdmunits$.  
Adopting this value does not qualitatively modify any 
of the above conclusions but does suggest there are
additional halos contributing modestly to DM$_{\rm FRB}$. 
Halo gas models with
lower $f_b$ imply larger $f_{\rm IGM}$ such that the
DM$_{\rm cosmic}$ values are nearly invariant.

The line of sight towards FRB~181112 also intersects the edge of the southern Fermi Bubble of our Galaxy \cite{Su2010}, as 
depicted in Fig.~\sfigFB. 
The Fermi Bubbles are plumes of magnetized plasma emanating 
from the Milky Way's Galactic Center \cite{Su2010}, and are 
known to host a large mass of cool and ionized
entrained material originating in the Milky Way's nuclear wind \cite{Bordoloi2017,Karim2018}. 
Thus, the DM along this direction will include an additional contribution 
from the Fermi Bubbles. 
Following \cite{Bordoloi2017}, we estimate the path length of FRB~181112 
through the Fermi Bubbles and estimate the DM contribution 
towards FRB~181112 as 
DM$_{\rm FB}  < 5 \left( \frac{n_e}{10^{-3} \rm cm^{-3}} \right)\,{\rm pc\,cm}^{-3}$. Even for a sightline that
passes through the center of the Fermi Bubbles, 
we predict only a moderate contribution of 
DM$_{\rm FB}  \sim 20 \left( \frac{n_e}{10^{-3} cm^{-3}} \right)\,{\rm pc\,cm}^{-3}$. 
Such a small contribution will be immeasurable in individual FRB
sightlines. 
Estimates for the magnetic field of the 
Fermi Bubbles \cite{carretti+13}
are sufficiently large to account for the 
observed rotation measure of FRB~181112.

\subsection{Rotation Measure and a limit on $B_\parallel$ for \fgname}
\label{sec:RM}

From the measured rotation measure of FRB~181112
(RM $= \rmvalue \pm \rmsig \, \rmunits$), we set an upper limit
to the organized magnetic field in \fgname.
In the following we adopt ${\rm RM} < \rmvalue \, \rmunits$
for the field associated with \fgname\ because our
Galaxy including the Fermi Bubbles ($\S$~\ref{sec:DM})
contributes to the measured signal.
The former is estimate to be
RM$_{\rm Galactic} = -2 \pm 11~\rmunits$ 
along the sightline to FRB~181112 \cite{oppermann+12}.

The equation relating RM and $B_\parallel$ is:

\begin{equation}
{\rm RM} = \frac{e^3}{2 \pi m_e^2 c^4} \int B_\parallel \, n_e \, ds / (1+z)^2
= 8.12 \times 10^2 \, {\rm rad \, m^{-2}} \,
\frac{B_\parallel}{(1+z)^2}  \int \, n_e \, ds \;\; ,
\label{eqn:RM}
\end{equation}
where the latter expression assumes a constant magnetic
field and the units for the quantities are
$\mu$G for $B_\parallel$, cm$^{-3}$ for $n_e$,
and kpc for $ds$.
For the density integral, we consider several models
for the gas in galactic halos.  Specifically,
we take the NFW, mNFW ($\alpha=2, y_0=2$),
\cite{mb04},
\cite{fielding+17}, and \cite{stern+18}
models described in \cite{xyz19}.
These span a range of assumptions on the fraction
of baryons retained within galactic halos
$f_b$ and the density profile $n_e(r)$.
Scaling to our estimated halo mass for \fgname\
of $\mmhalo = 10^{\mhalov} \mmsun$,
we calculate $\int n_e \, ds$
over an interval $\pm 15$\,kpc and $\pm r_{\rm vir}$
at $\mrperp = \rperpv$\,kpc.  
The former, corresponding to $L = 30$\,kpc (as assumed in the
main text), encompasses the majority of the gas and minimizes
assumptions related to the outer halo.
We also note that typical values for $n_e$ near the
effective mid-plane of these
models are $\approx 10^{-3} \, \rm cm^{-3}$ 
(also the fiducial
value adopted in the main text).
Inverting equation~\ref{eqn:RM}
we derive estimates for $B_\parallel^{\rm max}$,
the maximum allowed organized magnetic field
(ignoring any field reversals).
These are illustrated in Fig.~\sfigB\ for the
suite of models considered.
These values are lower than previous estimations
for halo gas \cite{bernet08} and far lower than 
measurements from regions above the plane of disks
in nearby galaxies (e.g.\ \cite{woodfinden19}).

\subsection{Scattering Analysis}
\label{sec:scattering}

At cosmological distances the characteristic temporal broadening time due to the multipath propagation caused by scattering in an inhomogeneous plasma is \cite{macquart13}
\begin{eqnarray}
\tau = \frac{1}{c k} \left( \frac{r_{\rm F}}{r_{\rm diff}} \right)^2 = \frac{D_L D_{LS} \lambda_0}{2 \pi \, c \, k \, D_S (1+z_L) r_{\rm diff}^2}.  \label{tauApprox}
\end{eqnarray}
with $D$ the angular diameter distance to the source (S),
screen (L), or the screen to the source (LS), 
$z_L$ is the redshift to the screen,
$r_{\rm diff}$ is the diffractive scale,
and where $k=2 \pi /\lambda_0$ and $\lambda_0$ is the wavelength in the observer frame.  For a power-law power spectrum of density inhomogeneities with index $\beta$ between an inner scale $l_0$ and outer scale $L_0$,  $r_{\rm diff}$ is the point at which the phase structure function\cite{macquart13}, 
 \begin{eqnarray}
D_\phi(r) &=& \begin{cases}
2^{4 - \beta} \pi^2 r_e^2 \, l_0^{\beta-4} \, {\rm SM}\, \frac{\lambda_0^2}{(1+z_L)^2}  \Gamma \left( 2 - \frac{\beta}{2} \right) r^2, & \quad {r< l_0}, \\ 
 \frac{2^{5 - \beta}}{\beta-2} \pi^2 \, r_e^2 \, {\rm SM} \, \frac{\lambda_0^2}{(1+z_L)^2} 
 \frac{\Gamma \left( 2 - \beta/2 \right)}{ \Gamma \left(\beta/2 \right)} r^{\beta-2},
& \quad {r \gg l_0},
\end{cases}
\label{eqn:Dphicases}
\end{eqnarray}
satisfies $D_\phi (r) = 1$.
Here, ${\rm SM} = \int dz\,C_N^2(z)$ is the scattering measure of the scattering medium (see, e.g. \cite{coles+87}),
$C_N$ is the amplitude of turbulence per unit length,
and $r_e$ is the classical electron radius.  The value $\beta=11/3$, corresponding to Kolmogorov turbulence, is often found to apply to turbulent plasma in the interstellar medium \cite{ars95}.

For scattering on a single thin plane, the pulse broadening kernel follows, to a good approximation, the function $P_D(t) = \exp (-t/\mdtpulse)$.  The pulse broadening time, \dtpulse, is observationally referenced to the time at which the pulse falls to $e^{-1}$ of its maximum value.  
\cite{LeeJokipii75} 
shows that this occurs approximately 
at $\mdtpulse \equiv (3/2) \tau$ (noting the factor of 1/2 difference between $\tau$ above and the 
definition of $t_D$ in \cite{LeeJokipii75}).   

For the geometry applicable to scattering by the foreground galaxy, one has $D_{\rm S} = 1.23\,$Gpc, $D_{\rm L} = 1.05\,$Gpc and $D_{\rm LS} = 0.26\,$Gpc.  For the scattering time to fall below $\mdtpulse< 40\,\upmu$s at 1300\,MHz along this line of sight requires a value of $r_{\rm diff} > 9.2 \times 10^{10}\,$cm in any inhomogeneous plasma located within the environs of the foreground galaxy.  

The diffractive scale, the length scale between adjacent ray paths over which the medium induces one radian of phase difference, is evaluated from the scattering timescale for the geometry relevant to \fgname; attributing all the scattering along this line of sight to this galaxy yields a minimum estimate of its scattering strength.  Assuming the diffractive scale to lie outside the inner scale of the turbulence (i.e. the dissipation scale; typical values are 200-1000\,km for Galactic interstellar turbulence), and that the density fluctuations follow a Kolmogorov spectrum, $\beta=11/3$, one has $r_{\rm diff} > 9.2 \times 10^{10}\,{\rm cm}$ and ${\rm SM} < 5.6 \times 10^{13}\,$m$^{-17/3}$.  We estimate the total variance in the column density (i.e. the variance in the dispersion measure) in the medium using the fact that the value of the phase structure function at the
outer scale of the turbulence is twice the phase variance, $D_\phi (L_0)  = 2 \langle \Delta \phi^2 \rangle$, and $\Delta \phi = r_e [\lambda_0/(1+z_L)] {\rm DM}$.  For the value of ${\rm SM}$ derived above the root mean square of the column density variations implied by the scattering is:
\begin{eqnarray}
\langle (\Delta {\rm DM})^2 \rangle^{1/2} < 0.089 \left( \frac{L_0}{1\,{\rm pc}} \right)^{5/6}\,{\rm pc\,cm}^{-3},
\end{eqnarray}
where have normalised to a fiducial value of $L_0=1\,$pc, noting that this outer scale is most plausibly also the scale at which the turbulence is driven.

The estimate of the column density variance is the most basic result derivable involving the fewest number of unknown parameters (to wit, the outer turbulent scale).  However, if the thickness of the medium is known one may obtain the variance in the in-situ electron density $n_e$. 
For a scattering medium thickness of $\Delta L$, in which the average value of $C_N^2(z)$ is uniform throughout the medium, one has $ C_N^2 < 3.7 \times 10^{-8} \left( \Delta L/50\,{\rm kpc} \right)^{-1} \,$m$^{-20/3}$.  
For density fluctuations whose variance on the outer scale is $\alpha^2 \langle n_e \rangle^2$, where $\langle n_e \rangle$ is the mean density,
one has (\cite{AnanNara1988}) $C_N^2 = 0.066 \alpha^2  \langle n_e \rangle^2 L_0^{-2/3}$.  For the value of $C_N^2$ derived here, the upper limit on the electron density is
\begin{eqnarray}
\langle n_e \rangle < 2.3 \times 10^{-4} \, \alpha^{-1} \left( \frac{L_0}{1\,{\rm pc}} \right)^{1/3} \left( \frac{\Delta L}{50\,{\rm kpc}} \right)^{-1/2} \, {\rm cm}^{-3}.
\end{eqnarray}

These limits apply irrespective of whether the baryons in the plasma of the foreground galaxy are smoothly distributed (where geometric optics would suffice to treat the problem) or whether they are inhomogeneous on scales smaller than the Fresnel scale (hence diffraction occurs).  The limit on $r_{\rm diff}$ is still well below the Fresnel scale of $r_{\rm F} = 5.1 \times 10^{13}\,$cm (i.e. the Fresnel scale exceeds the limit on the diffractive scale by a factor of 540, so diffractive theory is applicable here).  Indeed, irrespective of whether the radiation propagates through smooth lensing structures on scales above $r_{\rm F}$, the absence of diffractive scattering and temporal smearing in the medium nonetheless yields a valid upper limit on the root-mean-square column density variations in the medium, subject to the assumption that the density fluctuations follow a power law power spectrum.   

Were the phase to be smooth on scales comparable to $r_{\rm F}$ (i.e. for a lensing formulation to be readily applicable), the root mean square DM fluctuations would be less than $4.7 \times 10^{-4} \, (L_0/1\,{\rm pc})^{5/6}\,$pc\,cm$^{-3}$ and the limit on the in-situ electron density would be
\begin{eqnarray}
\langle n_e \rangle < 1.2 \times 10^{-6}  \alpha^{-1} \left( \frac{L_0}{1\,{\rm pc}} \right)^{1/3} \left( \frac{\Delta L}{50\,{\rm kpc}} \right)^{-1/2} \, {\rm cm}^{-3}.
\label{eqn:scatteringconstraint2}
\end{eqnarray}

\subsubsection{Interpretation of scattering constraint}
\label{sec:scatteringmotivations}

Our analysis of scattering is via equation~(\ref{eqn:scatteringconstraint2}).  This formula constrains the level of turbulence both in hot virialized gas and in colder $10^4$K gas.  Our measurement of the scatter broadening, $\tau$, also constrains the refractive scattering of a cool population 
of parsec-scale clouds as has been hypothesized \cite{mccourt+18}.  In what follows we discuss how our constraint pertains to these different scenarios.

\noindent
{\bf Turbulence in hot virialized gas}

  The canonical picture is that massive dark matter halos with 
  mass \mhalo\ are filled with virialized gas at virial temperature $T_{\rm vir}\approx 10^6 {\rm ~K~} [\mmhalo/(10^{12} M_\odot)]^{2/3} (1+z)$ with a filling factor of nearly unity.  This gas is stirred on kiloparsec scales by galactic winds, AGN-driven bubbles, and inflows, 
  with observations suggesting a mach number
  ${\cal M} \approx 1$ on $\sim 10~$kpc scales \cite{werk+14, mcquinn18}.  This stirring drives a turbulent cascade that, if the electron density fluctuations follow the spectrum of Kolmogorov turbulence to $r_{\rm diff}\sim 10^{10}$cm, leads to the scattering constraint represented by equation~(\ref{eqn:scatteringconstraint}).  Here we discuss whether these assumptions are likely to hold. 

There are multiple effects that can suppress the cascade on scales larger than $r_{\rm diff}\sim 10^{10}$cm \cite{lithwick01}.  
  The amplitude of density fluctuations responsible for scattering may be suppressed if the gas can cool for eddies with size $\ell > L_{\rm MHD}$, where $L_{\rm MHD} \sim L_0 \beta^{-3/2}$ is the scale where the magnetic energy density becomes larger than the turbulent kinetic energy and $\beta$ is the ratio of thermal to magnetic pressure. (Below $L_{\rm MHD}$ the isobaric component of these larger-scale density fluctuations are passively mixed by the incompressive shear Alfv\'en waves that characterize magnetohydrodynamic turbulence, generating small-scale density fluctuations \cite{1995ApJ...438..763G, lithwick01}.  If the turbulence is isothermal at $L_{\rm MHD}$, then there are no density fluctuations to mix.)   For virialized gas, the cooling time is expected to be $\gtrsim 10$ times longer than sound crossing time on the halo scale to avoid strong thermal instability \cite{mccourt12} and so the turbulence should be adiabatic.  The turbulent cascade could be terminated on sufficiently small scales by collisions with neutrals or ions \cite{lithwick01}.  However, we find that there are insufficient neutrals for the former damping mechanism to be effective (as the virialized CGM is highly collisionally ionized).
  For ion diffusion, eddies with turnover time equal to the proton diffusion time are damped \cite{lithwick01}, corresponding to 
  \begin{equation}
 \ell_{0}^{\rm pd} \sim 6 \times10^{13} {\rm ~cm~} \left( \frac{1~{\rm kpc}}{L_0}\right)^{1/2} \left( \frac{10^{-4} {\rm ~cm}^{-3}}{n_e/ \beta} \right)^{3/2}.
 \label{eqn:elldamp}
 \end{equation}
 The value $  \ell_{0}^{\rm pd} = 6 \times10^{13} {\rm ~cm} $ is
 several orders of magnitude larger than $r_{\rm diff}$ assumed by our constraint, weakening our constraint on $\langle n_e \rangle$ by $\sim (r_{\rm  diff}/\ell_0^{\rm pd})^{1/6}$ (eqn.~\ref{eqn:Dphicases}); this translates into a suppression by factor of a few.
 Furthermore, electron diffusion can damp fluctuations on a scale, which \cite{lithwick01} estimates will damp the density fluctuations by a factor of $\beta^{-1/2}$ on the somewhat larger scale of $\sqrt{m_p/m_e} ~\ell_{in}^{pd}$, where $m_p$ and $m_e$ are the electron and proton masses.

\noindent
{\bf Turbulence in $10^4$K gas}

 If not in the virialized phase, intrahalo gas is most likely to be in a denser, $T\sim 10^4$K photoionized phase.  Observations suggest volume filling fractions of $f_{V} \sim 10^{-3} -10^{-2}$ for halos of similar masses to \fgname \cite{werk+14, mccourt+18}, with the latter corresponding to a mass fraction of 0.5 in the cold phase.   Despite this phase likely having $f_{V} \ll 1$, its higher density and potentially smaller driving scale for turbulence may allow it to dominate the scattering.   We can re-write equation~(\ref{eqn:scatteringconstraint}) rescaled to values that are more relevant for this case:
\begin{equation}
 \left(\frac{\langle n_e \rangle}{0.1~{\rm cm}^{-3}}\right)  \left( \frac{f_{V} }{10^{-3}} \right)^{1/2} <  0.6  \alpha_{\rm rms}^{-1}   \left( \frac{\Delta L}{50\,{\rm kpc}} \right)^{-1/2} \left( \frac{L_0}{1\,{\rm kpc}} \right)^{1/3}  \left(\frac{\mdtpulse}{40 \, \rm \mu s} \right)^{5/12}  
\,{\rm cm}^{-3},
\label{eqn:coldcloud}
\end{equation}
where we have explicitly stated a volume filling fraction that was previously incorporated into our definition in our $\alpha$ so that $\alpha = \alpha_{\rm rms} f_V^{1/2}$.
The fiducial value of $\langle n_e \rangle = 10^{-1}$cm$^{-3}$ is based on pressure equilibrium with the virialized gas assuming $T=3\times 10^6$K (in accord with that expected for \fgname) and $n_{\rm e, vir} = 0.3\times 10^{-3}$cm$^{-3}$ -- roughly the minimum of the circumgalactic
medium (CGM) models considered in Fig.~\sfigB\  at 30~kpc.   This formula suggests $f_{V} \sim 10^{-3}$ for $L_0 \sim 1~$kpc.   The constraints are an order of magnitude stronger if we assume $L_0 \sim 1~$pc, as might be the case if thermal instability drives the turbulence \cite{mccourt+18}.
Our formula for the scattering timescale from Kolmogorov turbulence is 
nearly identical (with slight differences in normalization) to that for parsec-scale clouds presented in \cite{vp19}.  
Self-shielding of the metagalatic ionizing background should start to diminish the electron fraction for $\left(\frac{\langle n_e \rangle}{0.1~{\rm cm}^{-3}}\right)  \left( \frac{f_{V} }{10^{-3}} \frac{\Delta L}{50\,{\rm kpc}} \right)^{1/2}\gtrsim 1$ at $z=0.37$, somewhat larger than the values we are able to rule out for the fiducial numbers in equation~(\ref{eqn:coldcloud}).

Let us now consider whether the density fluctuations associated with the turbulent cascade may be suppressed in the cold $10^4$K clouds, which would weaken our constraints on $\langle  n_e \rangle$.
Following \cite{lithwick01}, below the cooling scale $L_c \equiv (c_s t_{\rm cool})^{3/2} L_0^{-1/2}$ where $c_s t_{\rm cool} \sim 1 ~n_e^{-1}$~pc \cite{mccourt+18} is the distance a sound wave travels with speed $c_s$ in a cooling time $t_{\rm cool}$, the turbulence is nearly isothermal, suppressing small-scale density fluctuations from passive mixing as described for the previous scenario.  Assuming that this is larger than $ L_{\rm MHD}$ (requiring $L_0 < 1 {\rm ~pc~} n_e^{-1} \beta^{3/2}$), density fluctuations will be instead sourced by either slow waves or passively mixing Reynolds stresses present at $\gtrsim L_{\rm MHD}$.  The density fluctuations from slow waves are suppressed by a factor of $\beta^{-1/2}$ relative to hydrodynamic scales; our result that the large-scale magnetic field is below equipartition suggests a suppression.  
For eddies 
with $\ell <L_{\rm MHD}$, density fluctuations sourced by Reynolds 
stresses are suppressed by the factor $\sqrt{c_s t_{\rm cool}/L_0}$ relative to the Kolmogorov extrapolation \cite{lithwick01}. 
Thus, for driving on parsec scales, suppression of the cascade may not be expected.  However, for kiloparsec-scale driving (for which Reynold stresses contribute negligibly to the small-scale density variations), the amplitude of density fluctuations in the cascade is not suppressed if $\beta \sim 1$ or if $L_c < L_{\rm MHD}$.  Otherwise a suppression with $\alpha \propto \beta^{-1/2}$ is expected.  We again expect proton diffusion to cutoff the cascade, but the higher densities of the cold gas make this cutoff less important than for the virialized gas, especially for kiloparsec driving scales (eqn.~\ref{eqn:elldamp}).  In conclusion, the parameter space for when the turbulent cascade in the cold gas is suppressed is complex.

\noindent
{\bf A refractive parsec-scale cloud population} 

Motivated by indications that there exists a population 
of cold parsec-scale clouds in the CGM \cite{mccourt+18}, we consider the scattering of spherical parsec-scale clouds that are not turbulent.  Non-turbulent clouds might occur if they survive quiescently for many sound crossing times after they are produced, as hypothesized in \cite{mccourt+18}.  (If turbulently driven on the cloud scale [with an undamped cascade to $r_{\rm diff}$] the scatter broadening yields the limit given by eqn.~\ref{eqn:coldcloud} with $L_0$ being the cloud size \cite{vp19}.)   Here we briefly sketch the derivation of refractive time delays for spherical clouds of size $R$.   While a spherical cloud is certainly an idealization, additional cloud structure will only increase the amount of scattering.  We also consider the generalisation
to a range of cloud sizes. 

 For a single cloud at $r\ll R$, we can expand the phase structure function in a power series
 $ D_\phi^{\rm cl}(r) = \phi_0^2 {r^2}/{R^2} +...$, where $\phi_0$ is a constant characteristic phase, using that the phase structure function must be zero at $r=0$ and an even function in $r$.  
 Calculating \dtpulse\ requires $D_\phi^{\rm cl}(r)$ only for $r \ll R$, so that retaining just the first term in the power series is a sufficient approximation.  For spatially uncorrelated clouds with an average of $N$ intersected across the halo, the full structure function is $D_\phi(r) =  N D_\phi^{\rm cl}(r)$.  The diffractive scale is defined as $D_\phi(r_{\rm diff})/2 =1$ such that $r_{\rm diff} = \sqrt{2 /N} R/\phi_0$. 
  When we take $\phi_0$ to be the phase across the center of the cloud, the resulting expression for the scattering angle from this $r_{\rm diff}$ is identical to that calculated for refractive clouds (\cite{draine}; section~11.4), using a different method.  A numerical calculation shows that $\phi_0$ is $2.2$ times larger than the phase across the cloud center. 
  We used Monte-Carlo integration to verify our form for $D_\phi$ and measure its normalization.  The extreme curvature at the edge of our idealization of perfectly spherical, tophat clouds means that the integral to compute $\phi_0$ needs to be regulated to yield a convergent answer.  We find that $D_\phi$ changes little if we do not include sightlines that fall with rarity of a part in $10^4$ or instead $10^5$ from the cloud edge, much larger than the $N$ expected physically in the picture of \cite{mccourt+18}.

The scatter broadening time \dtpulse\ can be calculated from $r_{\rm diff}$ using equation~(\ref{tauApprox});  it yields a smaller \dtpulse\ than the case where the clouds are fully turbulent (eqn.~\ref{eqn:coldcloud}) and that reported in \cite{vp19} whose results we interpret as being appropriate for fully turbulent clouds.  However, before we interpret this refractive delay as scatter broadening, we must check whether multi-path propagation is expected.  There would only be multi-path propagation if the size of the scattering surface $\chi$ is $\gtrsim R$ so that an  independent set of structures is intersected, where 
$\chi^2  = 2 c \mdtpulse D_L D_{LS}/[D_{S} (1+z_L)]$.  For the angular diameter distances and redshift of \fgname, $\chi = 0.011{\rm pc}\times \sqrt{{\mdtpulse}/(40 \,\upmu s)}$, smaller than the parsec estimate for circumgalactic cloud sizes \cite{mccourt+18}.  Taking this criterion into account, our constraint on the refractive scattering of spherical clouds is
\begin{eqnarray}
\left(\frac{n_e}{0.1~{\rm cm}^{-3}}\right)\left( \frac{\Delta L}{50 {\rm kpc}} \frac{f_{V}}{10^{-3}} \frac{0.1 {\rm pc}}{R} \right)^{1/2}    <  0.2 \left(\frac{\mdtpulse}{40 \upmu s} \right)^{1/2} &{\rm if} &R \lesssim 0.011{\rm pc} \left( \frac{\mdtpulse}{40 \upmu s} \right)^{1/2}; \nonumber\\
\left(\frac{n_e}{0.1~{\rm cm}^{-3}}\right)\left( \frac{\Delta L}{50 {\rm kpc}} \frac{f_{V}}{10^{-3}} \right)^{1/2} \left( \frac{0.1 {\rm pc}}{R} \right)^{3/2}  <  1.6  &{~\rm otherwise,}&
\end{eqnarray}
although again self-shielding also limits the maximum $n_e^2 f_V \Delta L$ probed by scattering (see above).
The former equation applies for bubbles whose size is small enough to generate multiple extrema in phase, and the latter weaker constraint applies to larger bubbles (as we require a higher column to result in scatter broadening from multipath propogation).   While weaker than the model in \cite{vp19}, which assumed a structure function with the scaling of Kolmogorov turbulence, these constraints show that even a tiny volume filling factor of $f_{V} \sim 10^{-3}$ for $\sim0.1~$pc clouds -- about the scale anticipated in \cite{mccourt+18} -- is ruled out in \fgname.  While we have made the assumption of a single cloud radius, the radius only enters our calculation for \dtpulse\ via its dependence on $N$, the number of clouds intersected, as long as the clouds are sufficiently small to contribute to multipath propagation.  

\subsubsection{Other galaxy halo constraints from scattering}

Here we compare the limits obtained in this section with those deduced for other galaxies.

\noindent
{\bf M31}: The properties of the baryonic halo of M31 are constrained by angular broadening measurements of distant radio galaxies in the background of this galaxy obtained using widefield very long baseline interferometry (VLBI) at a 1640\,MHz \cite{Morganetal2013}.  These observations detect angular broadening of $\approx 15\,$\,milliarcseconds 
towards the centre of the galaxy, at a projected distance of 0.25\,kpc, which is attributed to an extreme scattering environment close to the centre of this galaxy.  However, some background radio sources at separations between 15 and 30 arcmin of the galaxy center possess angular sizes $<1\,$mas, thus placing an upper limit on the ionized baryons present both in the disk and halo of M31 through which the background radiation propagates (\cite{Morganetal2013}, their Table 3 and figure~7).  
(Other sources at similar projected angular distances are 
resolved but these are attributed to the finite angular sizes of the sources themselves.)

The absence of angular broadening at the 1\,mas level immediately places a lower limit on the value of $r_{\rm diff}$ associated with the scattering in the disk+halo of M31 of $7.2 \times 10^6\,$m.  This in turn implies an upper limit on SM of $1.6 \times 10^{17}\,$m$^{-17/3}$ under the assumption that $r_{\rm diff}$ exceeds the inner scale of the turbulence: no measurement of the frequency dependence of the (absent) angular broadening is available to verify this assumption, however this assumption is supported by measurements of interstellar turbulence in the Milky Way (e.g. \cite{SpanglerGwinn1990}).  This upper limit in turn implies a limit on the root-mean-square electron column which, expressed in terms of the usual units of dispersion measure, gives
\begin{eqnarray}
\langle \Delta {\rm DM}^2 \rangle^{1/2} < 4.7 \left( \frac{L_0}{1\,{\rm pc}} \right)^{5/6} \, \hbox{pc\,cm}^{-3}.
\end{eqnarray}
This is an upper limit on the halo baryon contribution in several senses: both because no angular broadening is detected and because any detection would include a contribution from both the disk and the halo of this galaxy.  

\medskip

\noindent
{\bf The Milky Way}: Measurements of (or limits on) the angular broadening or pulse broadening from sources located beyond the Milky Way constrain the properties of its ionized baryonic halo.  In general, temporal smearing limits are more constraining at low frequencies due to the 
$\mdtpulse \propto \nu^{-4}$ dependence of the scattering. The most stringent limit is due to the pulse smearing measurement of FRB\,170827 \cite{Farahetal2018}, from which a smearing time of $4.1 \pm 2.7\,\upmu$s at 842\,MHz is deduced from its $\approx 30\,\upmu$s-duration pulse structure \cite{Farahetal2018}.  The decorrelation bandwidth of the burst structure is 1.5\,MHz, which would imply a more stringent limit of $0.8\,\mu$s on the scattering timescale, but the origin and interpretation of the spectral structure in FRBs is presently less straightforward, so we conservatively use the limit deduced directly from the temporal structure of the FRB itself.  

The most conservative limit on the properties of the baryonic halo are derived by attributing all of the observed temporal smearing to our Galaxy (rather than any intermediate system or host galaxy).  For a fiducial effective screen distance of $d_{\rm screen} = 10\,d_{10}\,$kpc from Earth one derives a diffractive scale length $r_{\rm diff} = 3.5 \times 10^7\,d_{10}^{1/2}$m and ${\rm SM} = 7.6 \times 10^{14}\,d_{10}^{-5/6}$m$^{-17/3}$ and 
\begin{eqnarray}
\langle \Delta {\rm DM}^2 \rangle^{1/2} < 0.65 \left( \frac{L_0}{1\,{\rm pc}} \right)^{5/6} \left( \frac{d_{\rm screen}}{10\,{\rm kpc}} \right)^{-5/12} \, \hbox{pc\,cm}^{-3}.
\end{eqnarray}


\clearpage

\begin{figure}[h]
\includegraphics[width=\linewidth]{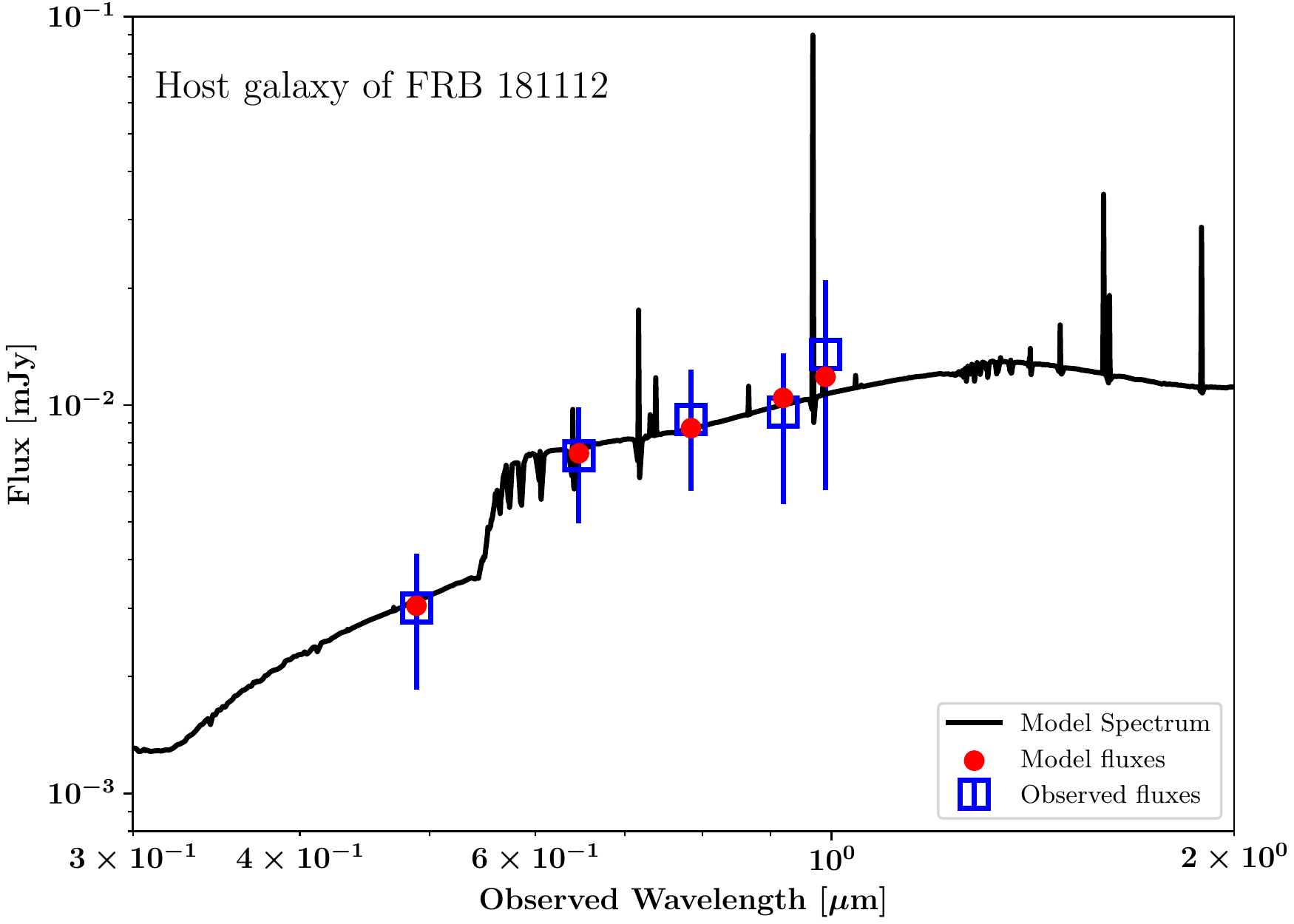}
    {\bf Fig.~\sfighostcigale:
    Spectral energy distribution of the host galaxy of
    FRB~181112.}  
    Comparison of observed
    fluxes of the host galaxy of FRB~181112 (blue squares) 
    against the best-fit model values
    (red circles) as derived with the CIGALE software package.
    The black curve shows the model spectrum.
    The derived properties (Table~\stabhost)
    include stellar mass
    $M_* = 10^{\logmstar} \, \mmsun$, reddening
    $E(B-V) = \fgcebv$\,mag, and SFR~$\approx 0.4 \pm 0.6 \; \mmsun \, \rm yr^{-1}$.
\end{figure}

\begin{figure}[h]
\includegraphics[width=\linewidth]{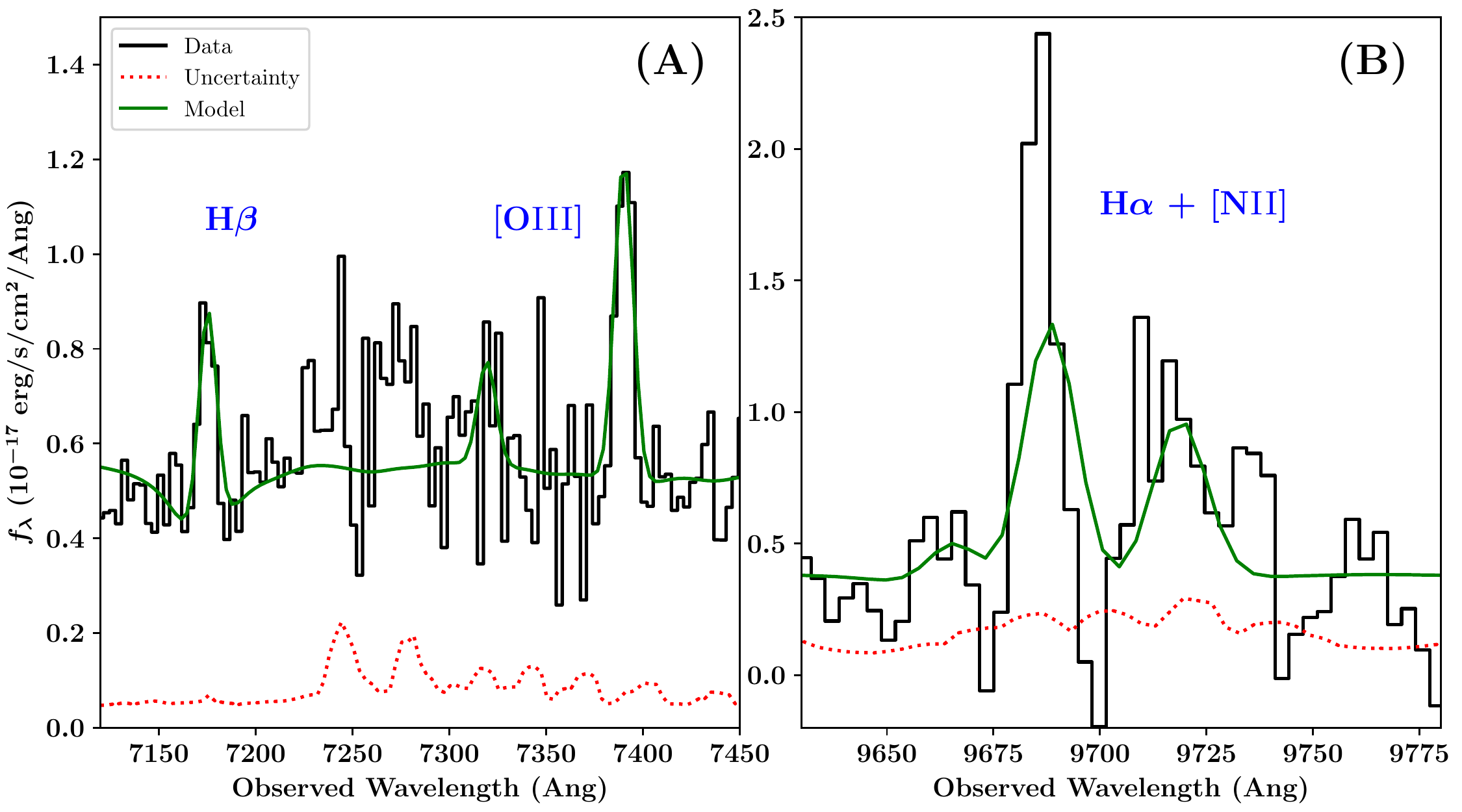}
    {\bf Fig. \sfighostppxf: Fitted profiles to spectral 
    features for the host galaxy of FRB~181112.}  VLT/FORS2 spectrum (specific flux $f_\lambda$; black histogram)
    and uncertainty (red dotted line) of several 
    nebular lines from the host
    galaxy of FRB~181112.  The
    best-fitting model (green solid line)
    was generated with the pPXF software package,
    which includes estimates for Balmer absorption.
    The data at $\lambda \approx 9700$\AA\ 
    suffer from substantial telluric absorption which has
    not been fully corrected.
\end{figure}

\begin{figure}[h]
\includegraphics[width=\linewidth]{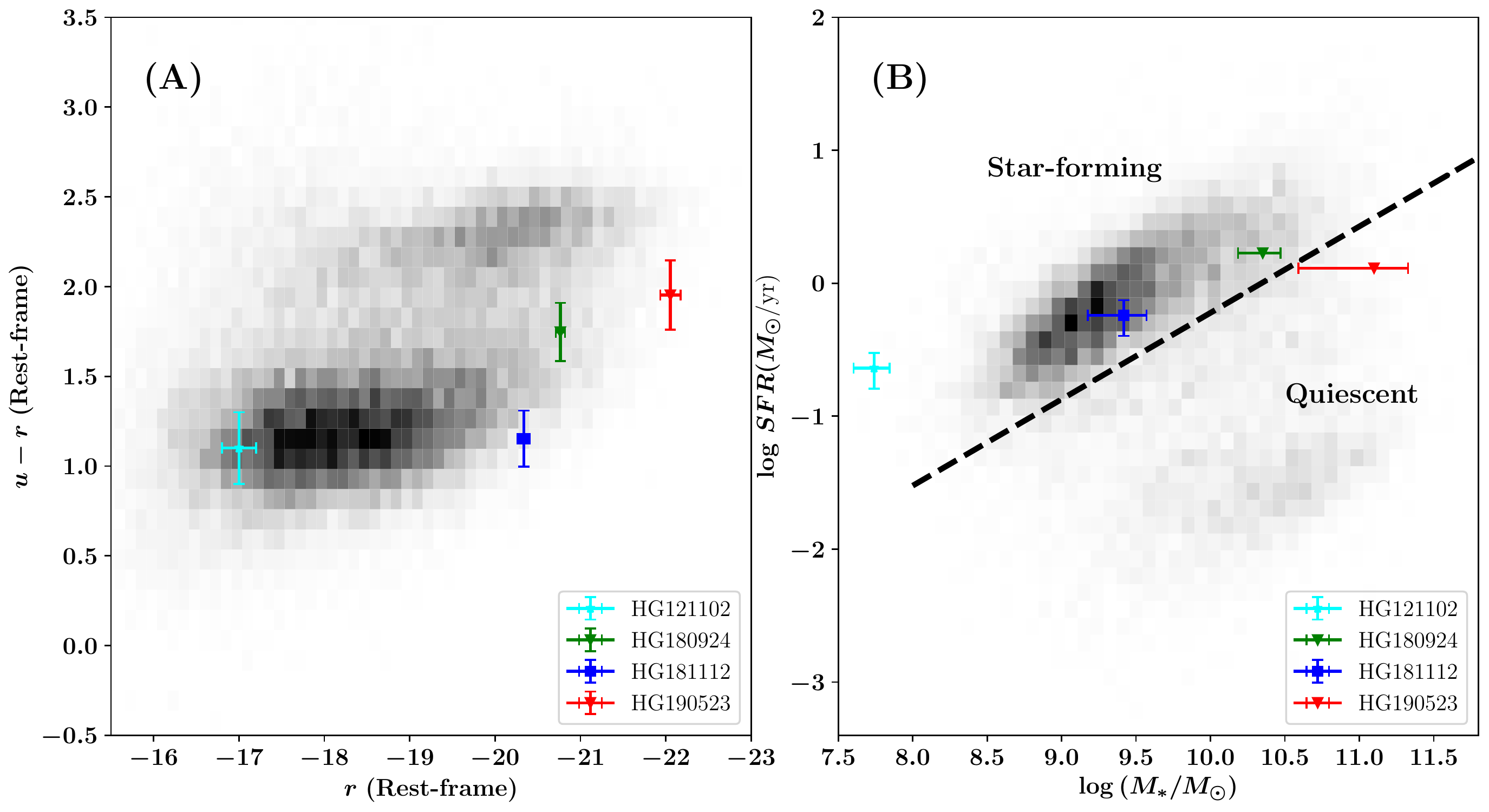}
    {\bf Fig. \sfighosts:  Comparison of FRB
    host galaxies with galaxies from the PRIMUS
    survey \cite{primus}}.
    Panel~A shows a rest-frame color-magnitude diagram comparing
    galaxy color with luminosity.  The PRIMIUS sample 
    (gray histogram)
    shows two main sequences of blue ($u-r \approx 1.1$\,mag) 
    and red ($u-r \approx 2.4$\,mag) galaxies.
    Panel~B compares galaxy SFRs with stellar mass with the 
    dashed line separating star-forming galaxies
    from quiescent galaxies.
    While the host of the Repeater FRB
    (labelled HG~121102; \cite{Tendulkar17}) occurred in a very faint, 
    blue, and star-forming
    galaxy, the non-repeating FRBs localized by ASKAP (labelled
    HG~180924 \cite{Bannisteretal2019}
    and HG~181112 \cite{supp}; thus far non-repeaters)
    and the Deep Synoptic Array (HG~190523; \cite{Ravietal2019}),
    have more luminous hosts (panel~A).
    These galaxies also have higher stellar mass (panel~B)
    and show a diversity of SFRs
    (the downward triangles for HG~180924 and
    HG~190523 indicate upper limits to the SFR).
\end{figure}

\begin{figure}[h]
    \includegraphics[width=\linewidth]{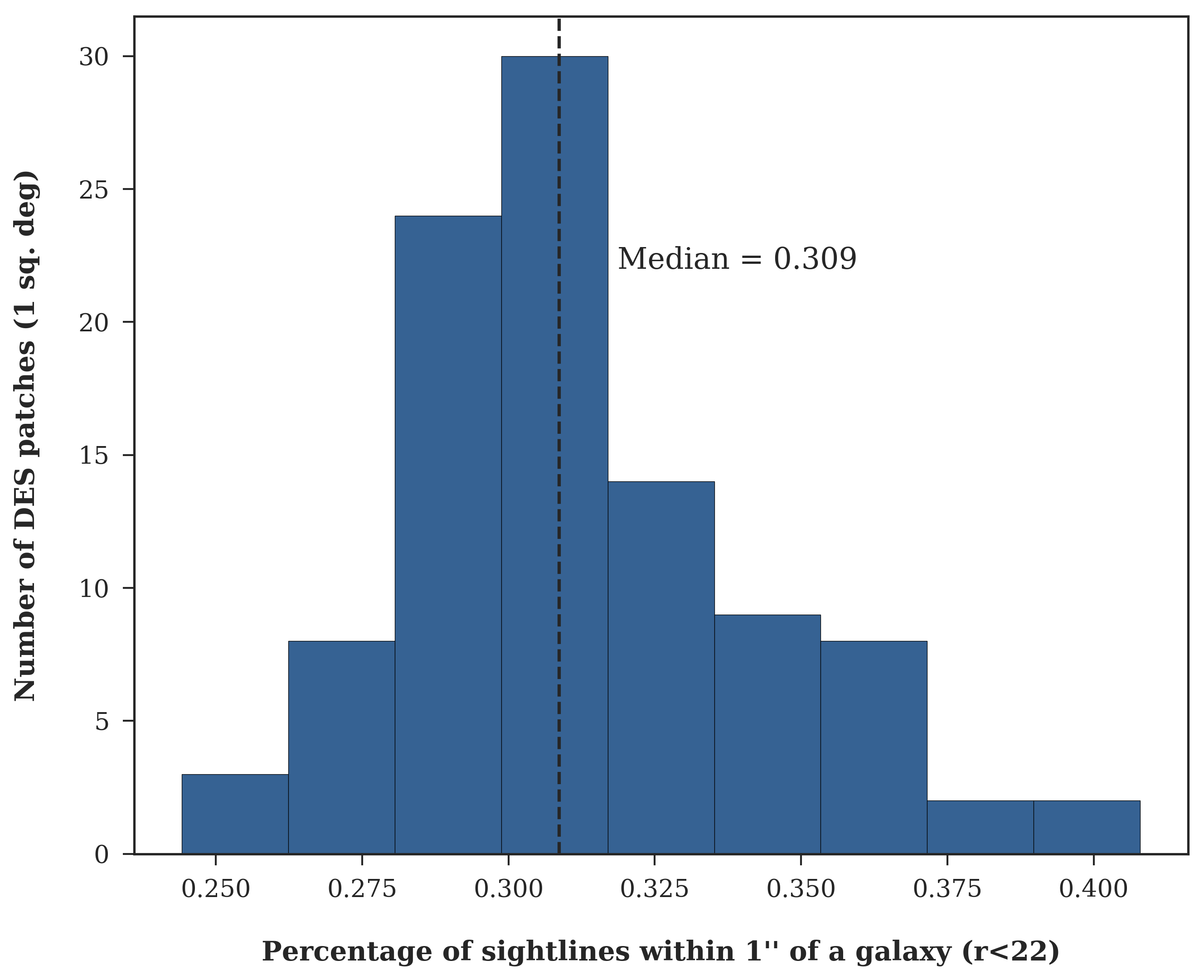}
    {\bf Fig. \sfigrandom:  Probability of a random association
    with a $r < 22$\,mag galaxy.
    }  
    The distribution from 100 trials of the percentage of 
    random sightlines falling within $1''$ of a galaxy with 
    $r<22$\,mag in a random 1 sq. deg patch of sky in the 
    DES-DR1 suvey. 
    The dashed black line represents the median value (0.3\%)
    of the full trial set.
    \label{fig:randAssoc}
\end{figure}

\begin{figure}[h]
    \includegraphics[width=\linewidth]{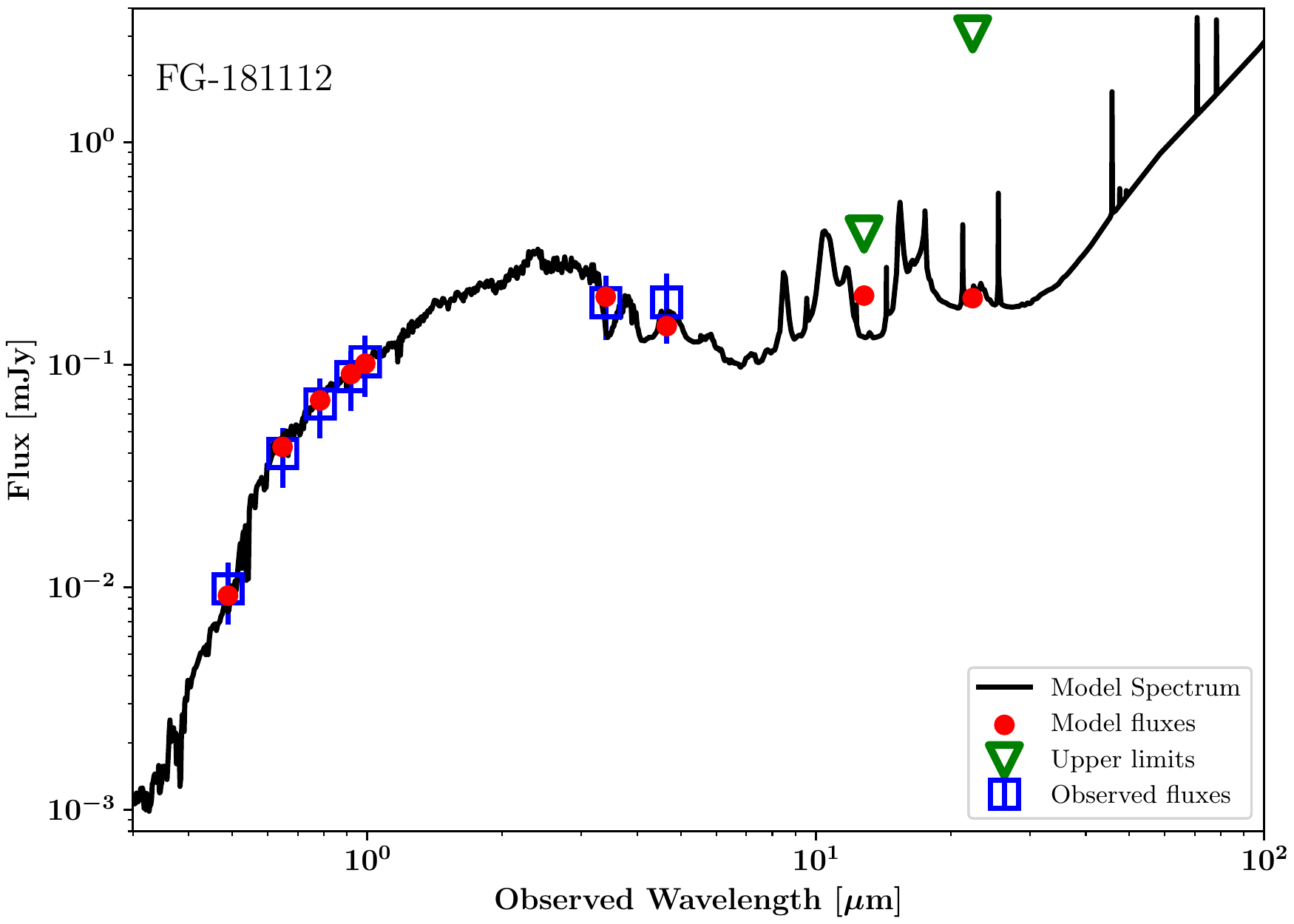}
    {\bf Fig. \sfigfgcigale:  
    Spectral energy distribution of the foreground galaxy \fgname.}
    Comparison of observed
    fluxes of \fgname\ (blue squares) 
    against the best-fit model values
    (red circles) as derived with the CIGALE software package.
    Green triangles represent upper limits on the 
    observed fluxes.
    The black curve shows the model spectrum.
    The derived properties include stellar mass
    $M_* = 10^{\logmstar} \, \mmsun$, reddening
    $E(B-V) = \fgcebv$\,mag, and AGN fraction $f_{\rm AGN} = 0.1$.
\end{figure}
    
\begin{figure}[h]
    \includegraphics[width=\linewidth]{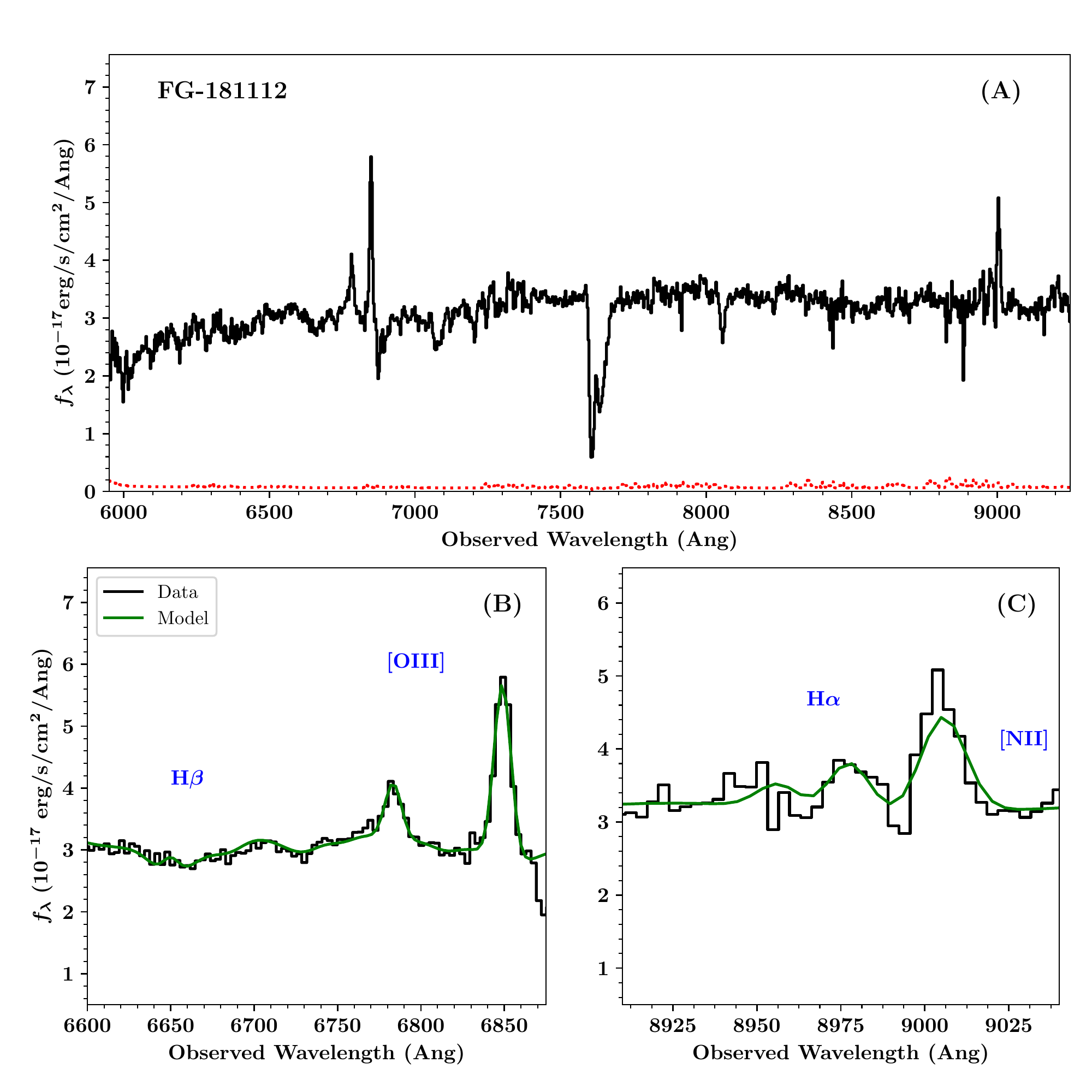}
    {\bf Fig. \sfigfgspec: Optical spectrum of the foreground
    galaxy \fgname.}
    (A) VLT/FORS2 spectrum of \fgname\ (black histogram)
    and estimated uncertainty (red dotted line).  The strong
    absorption feature at $\approx 7600$\AA\ is telluric.
    (B,C)  Zoom-ins on key nebular emission lines
    fitted using the pPXF software package (green solid line).
\end{figure}

\begin{figure}[h]
    \includegraphics[width=\linewidth]{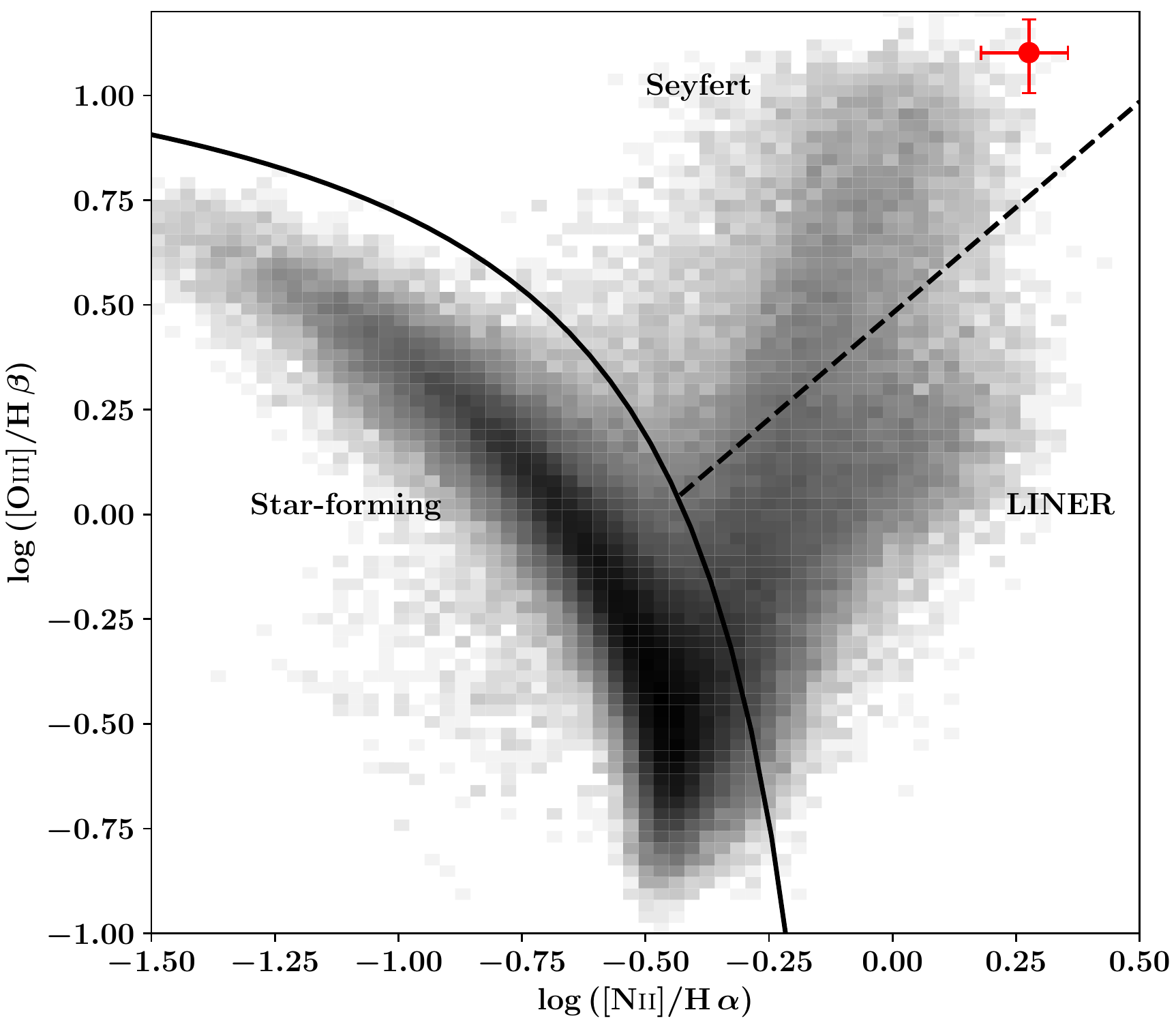}
    {\bf Fig. \sfigbpt:}
     Diagnostic plot for characterizing emission line galaxies. 
        Grey points show 
    emission line ratios from nearby galaxies ($z=0.02-0.4$) in the Sloan Digital Sky Survey, 
    restricted to have signal-to-noise greater than 5.
    The intensity scaling is logarithmic
    to accentuate regions away from the dominant, star-forming
    locus.   
    Black lines separate star-forming galaxies
    (solid; \cite{kewley01}) 
    from sources dominated
    by harder ionizing spectra (\cite{kauffmann03}),
    and the
    dashed line separates
    sources designated as Active Galactic Nuclei from 
    Seyfert and low-ionization nuclear emission-line region (LINER)
    galaxies \cite{fernandes10}.
    The line ratios of \fgname\ (red point)
    show it is a  Seyfert galaxy, i.e., it hosts 
    an active galactic nucleus.  
\end{figure}

\begin{figure}
    \includegraphics[width=\textwidth]{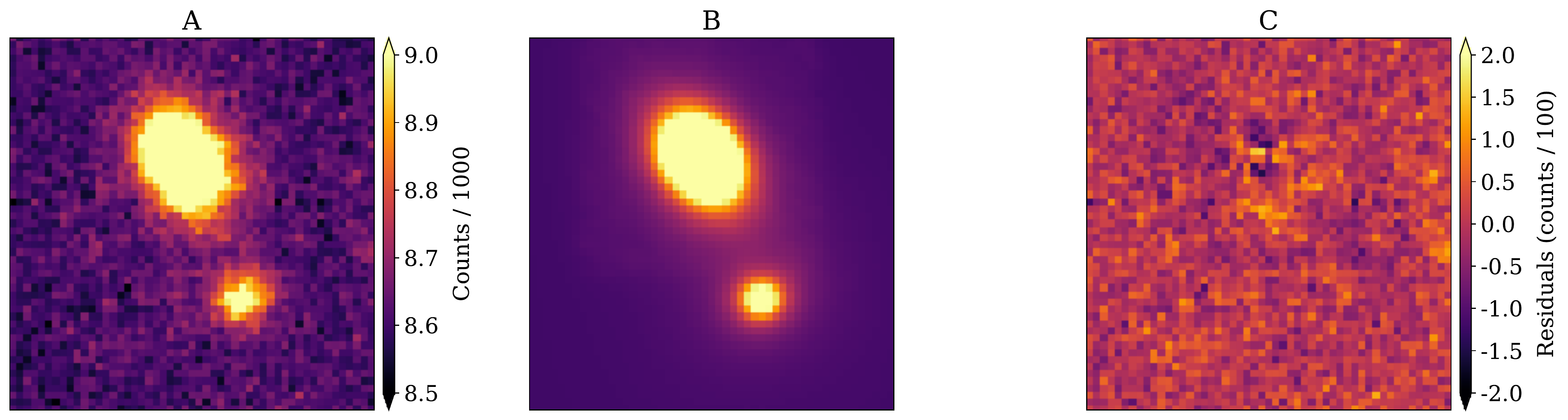}
    {{\bf Fig. \sfigsersic: Morphological analysis of the host
    and foreground galaxies for FRB~181112.}
    (A) FORS2 $I$-band image of the host (bottom) and foreground (top) galaxies. 
    The image is centered on the coordinates RA, Decl.\ 
    $\rm 21h49m23.8s, -52^\circ 58'12.2''$ (J2000), 
    has an angular extent of $12.6''$ on a side, and is oriented
    with N up and E to the left.
    (B) Best fit model with three components: two S\'{e}rsic profiles 
    and a constant sky background. The color scale is the same as (A).
    (C) The residuals between the data and the model.  
    }
    \label{fig:sersic}
\end{figure}

\begin{figure}[h]
\includegraphics[width=\linewidth]{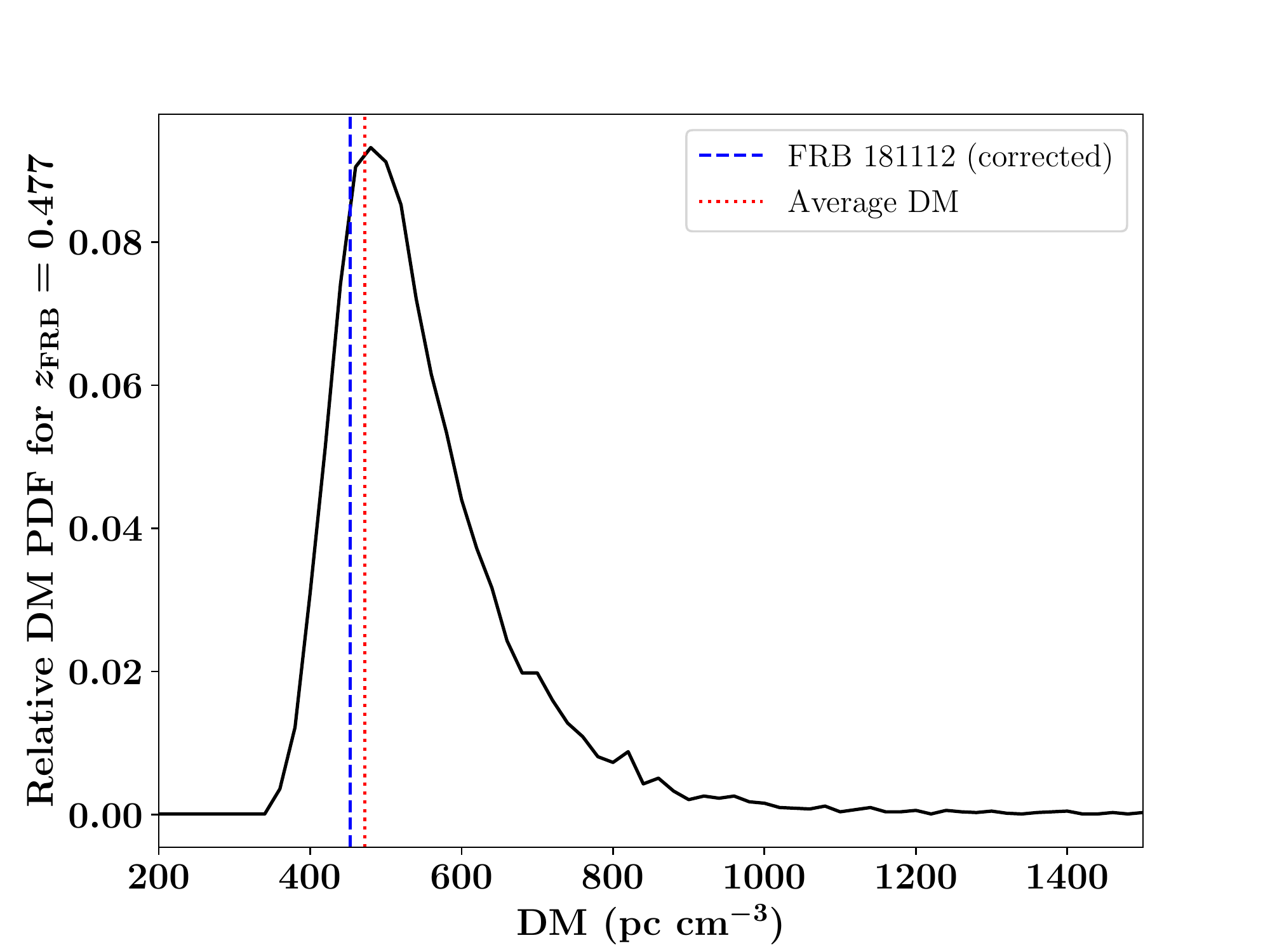}
    {\bf Fig. \sfigDMpdf:  Comparison of the corrected
    DM of FRB~181112 with random expectation.}
    The black curve is the estimated PDF for a random sample of FRBs
    at $z = \zfrb$ using the methodology described in \cite{xyz19}.
    These DM values do not include contributions from our Galaxy
    nor the host galaxy.
    The blue dashed vertical line indicates the DM value of FRB~181112
    corrected for the Galactic ISM, Galactic halo, and an assumed
    value for the host galaxy (see text).
    We find that it lies near the average expected value 
    for an FRB at this redshift (red, dotted line), 
    even though the sightline intersected the halo of \fgname.
\end{figure}

\begin{figure}[h]
\includegraphics[width=\linewidth]{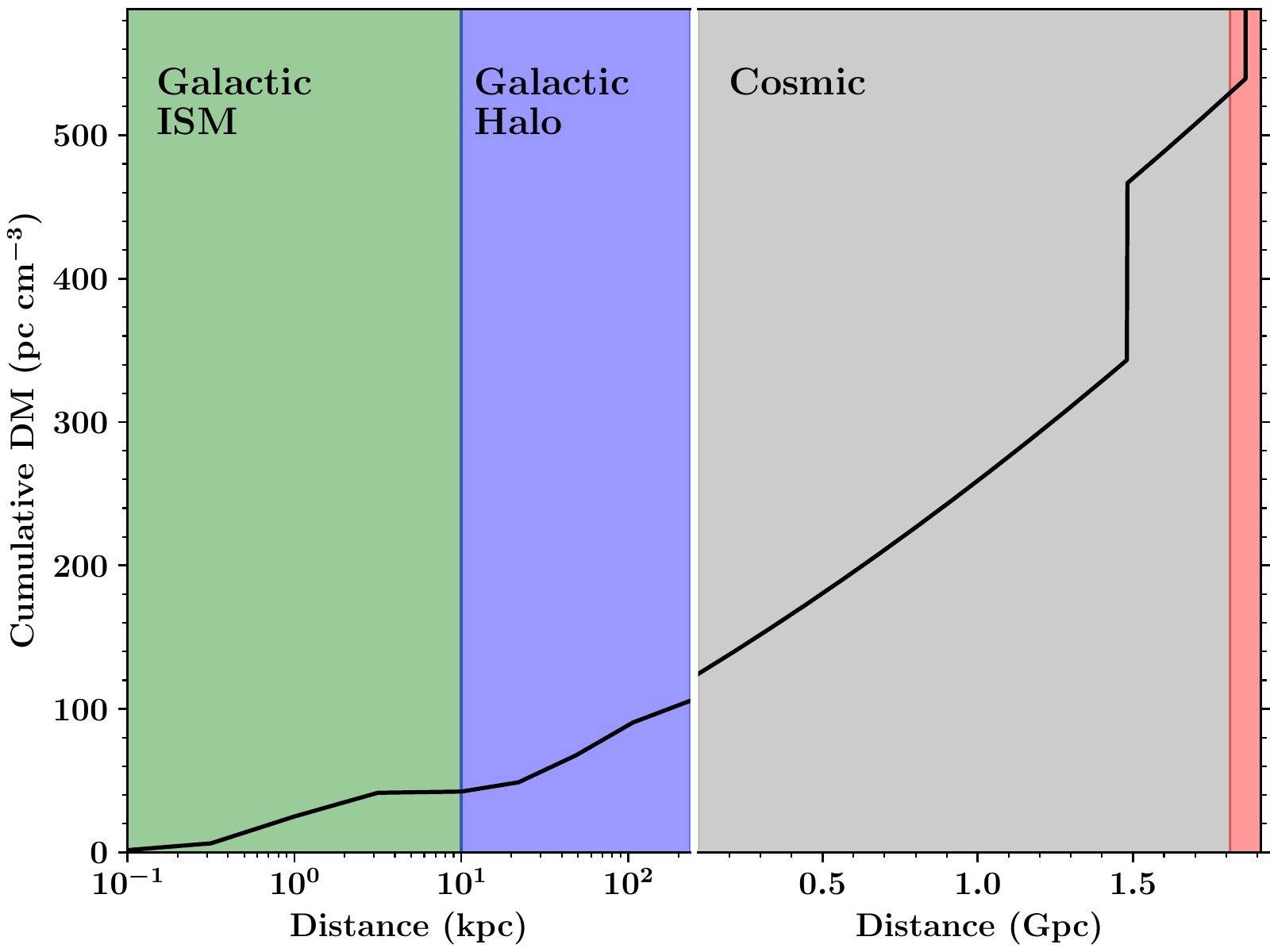}
    {\bf Fig. \sfigDMcum:  Diagram of the cumulative
    DM from Earth to FRB~181112.}  The cumulative DM 
    (black curve)  includes contributions from our
    Galactic ISM (green), Galactic halo (blue),
    a restricted cosmic contribution (gray, see text),
    and the host galaxy (pink).
    The jump at approximately 1.5~Gpc
    is due to the intersection with \fgname's halo.
    The y-axis of the figure terminates at the measured
    DM$_{\rm FRB} = \dmval \, \mdmunits$ for FRB~181112.  
    The curve shows that the FRB~181112 sightline is unlikely
    to have intersected many additional, massive halos.
\end{figure}

\begin{figure}[h]
\includegraphics[width=\linewidth]{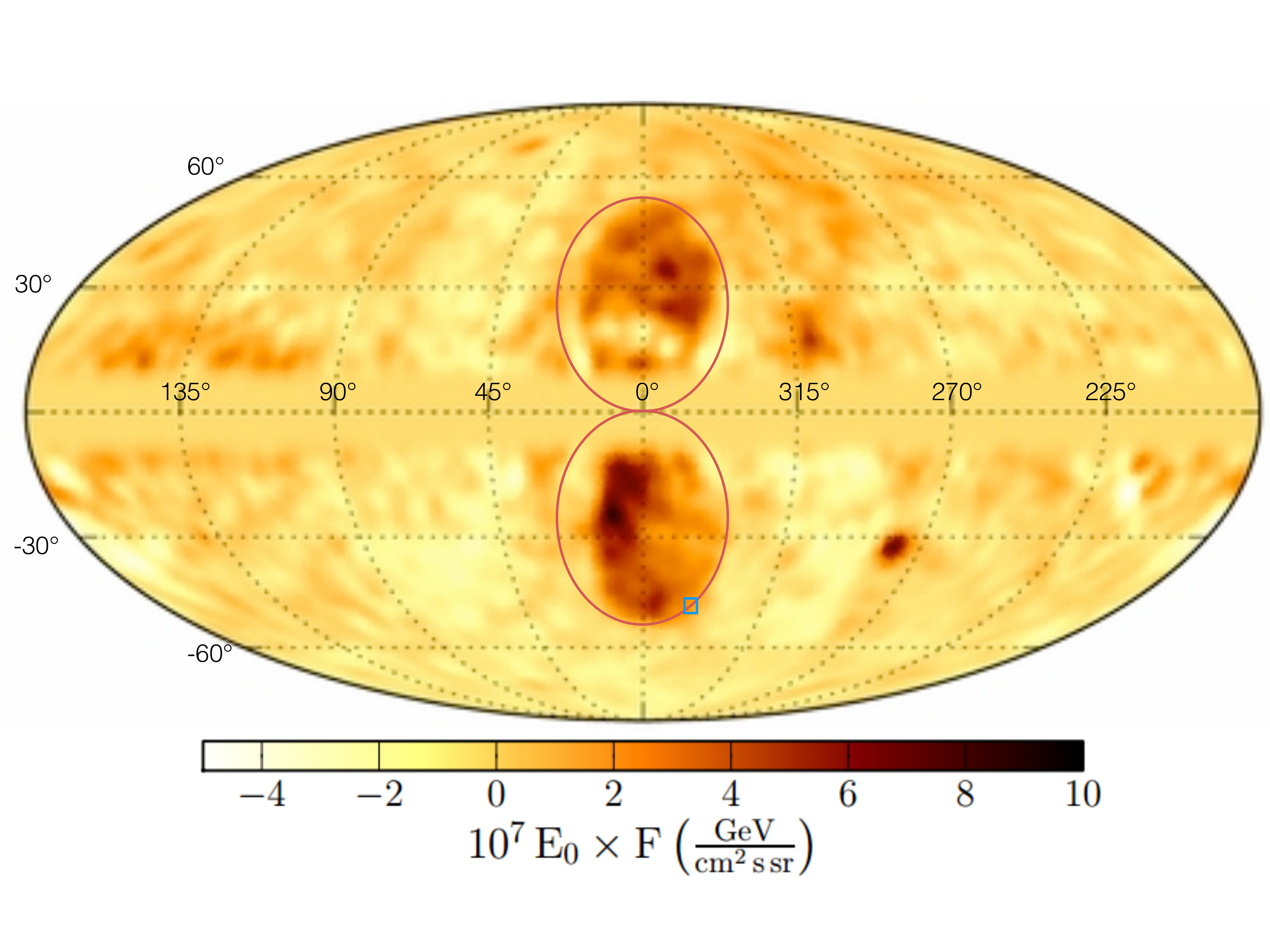}
    {\bf Fig. \sfigFB: The sightline to FRB~181112 intersects
    the southern Fermi bubble.}  
     The all-sky Fermi residual gamma-ray intensity map observed in the 3--10 GeV range, in Galactic coordinates centered on the Galactic Center (adapted from \cite{Ackermann2014,Bordoloi2017}). The 
     twin Fermi Bubbles are marked by the red ellipses.
     The blue square marks the line of sight through the Fermi Bubbles towards FRB~181112, which intersects the edge of the 
     southern bubble.
\end{figure}

\begin{figure}[h] 
\includegraphics[width=\linewidth]{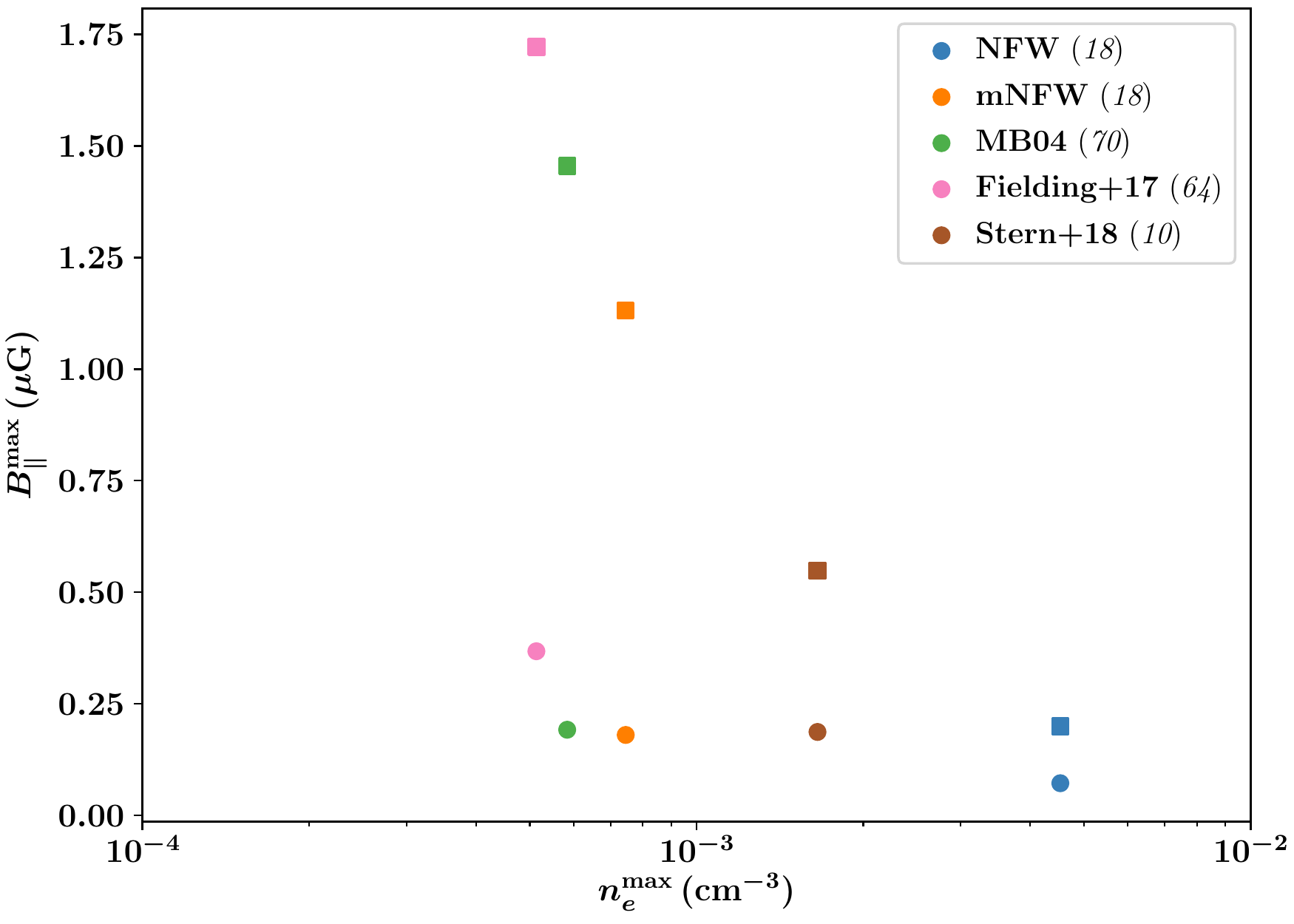}
    {\bf Fig. \sfigB:  
    Upper limits on the organized parallel magnetic field \bmax\ 
    in the halo of \fgname\ for a series of assumed
    CGM models.} 
    The circles (squares) integrate the
    rotation measure through the entire halo (or $\pm 15$\, kpc
    from the mid-plane).
    See $\S$~\ref{sec:RM} and \cite{xyz19} for details on the
    various CGM models shown.  
    We conclude that the organized magnetic field along the
    FRB sightline in the halo of \fgname\ 
    is less than $1 \, \upmu$G and likely less than $0.3 \, \upmu$G.
\end{figure}


\clearpage


\section*{Supplementary Tables}

\begin{table}[b]
\centering
\caption{{\bf Fitted positions and flux densities of the 3 brightest ASKAP field sources \label{tab:ASKAPsrcs}}
}
\begin{minipage}{270mm} 
\begin{tabular}{*{7}{c}}
\hline 
Source & RA & RA$_{\textrm{uncertainty}}$ & Dec. & Dec.$_{\textrm{uncertainty}}$ & S & S$_{\textrm{uncertainty}}$ \\ 
& (J2000) & (mas) & (J2000) & (mas) & (mJy) & (mJy beam$^{-1}$)
\\ 
\hline 
SUMSS 215155$-$531646 & 21:51:55.37 & 2100 & $-$53:16:47.5 & 1400.0 & 40.9 & 3.8 \\ 
SUMSS 215036$-$525538 & 21:50:37.16 & 2100 & $-$52:55:40.0 & 1400.0 & 33.9 & 3.8 \\ 
SUMSS 215606$-$532916 & 21:56:06.32 & 2900 & $-$53:29:11.2 & 1800.0 & 29.5 & 3.7 \\ 
\hline 
\end{tabular} 
\end{minipage} 
\end{table} 

\clearpage


\begin{table*}
\centering
\caption{{\bf Phase calibrators used for the ATCA observation.~The position of SUMSS~215155$-$531646 was derived using these calibrators. \label{tab:ATCAcal}}
}
\begin{minipage}{270mm} 
\begin{tabular}{cccc}
\hline 
Calibrator & Source & RA  & Dec.  
\\
(J2000) & (J2000)  & (J2000) & (J2000)
\\ 
\hline 

& \textbf{Calibrators used}   & & \\
& PKS B2117$-$642 & 21:21:55.0217  & $-$64:04:30.042  \\ 
& PKS B2215$-$508 & 22:18:19.0247  & $-$50:38:41.731  \\
\\
& \textbf{FRB~181112 field source}  & & \\
PKS B2117$-$642 & SUMSS 215155$-$53164 & 21:51:55.307
 & $-$53:16:46.56
 \\
PKS B2215$-$508 & SUMSS 215155$-$53164 & 21:51:55.291
 & $-$53:16:47.09 
  \\
\hline
\end{tabular}
\end{minipage}
\end{table*}

\clearpage
\begin{table}
\centering
\caption{{\bf Galaxy Photometry.}
DES-DR1 photometry were published in \cite{desdr1}
and WISE photometry were taken from \cite{Wright10}.
Uncertainty values of $-999.00$ indicate upper limits. 
\label{tab:photom}}
\begin{minipage}{270mm} 
\begin{tabular}{ccccc}
\hline 
Galaxy & Survey & Filter & Value (mag) & Uncertainty (mag)
\\ 
\hline 
Host & DES-DR1 & g & 22.71 & 0.09 \\ 
Host & DES-DR1 & r & 21.73 & 0.05 \\ 
Host & DES-DR1 & i & 21.49 & 0.06 \\ 
Host & DES-DR1 & z & 21.45 & 0.11 \\ 
Host & DES-DR1 & Y & 21.07 & 0.17 \\ 
Host & VLT & g & 22.57 & 0.04 \\ 
Host & VLT & I & 21.51 & 0.04 \\ 
\hline 
Foreground & DES-DR1 & g & 21.42 & 0.03 \\ 
Foreground & DES-DR1 & r & 19.90 & 0.01 \\ 
Foreground & DES-DR1 & i & 19.34 & 0.01 \\ 
Foreground & DES-DR1 & z & 19.03 & 0.01 \\ 
Foreground & DES-DR1 & Y & 18.86 & 0.02 \\ 
Foreground & VLT & g & 21.20 & 0.04 \\ 
Foreground & VLT & I & 19.20 & 0.02 \\ 
Foreground & WISE & W1 & 15.53 & 0.04 \\ 
Foreground & WISE & W2 & 14.89 & 0.06 \\ 
Foreground & WISE & W3 & 12.28 & -999.00 \\ 
Foreground & WISE & W4 & 8.59 & -999.00 \\ 
\hline 
\end{tabular} 
\end{minipage} 
\end{table}

\clearpage
\begin{table}
\centering
\caption{{\bf 
Physical properties of the host galaxy 
of FRB~181112 derived from CIGALE and nebular line analysis \label{tab:host}}
}
\begin{minipage}{270mm} 
\begin{tabular}{ccccc}
\hline 
Package & Parameter & Value & Uncertainty & Unit 
\\ 
\hline 
CIGALE & Stellar Mass & 0.26 & 0.1& $10^{10} M_\odot$ \\ 
CIGALE & AGN Fraction & 0.0 & 0.0 & \\ 
CIGALE & Rest-frame $u-r$ & 1.2 & 0.2 & mag  \\ 
CIGALE & Rest-frame $M_r$ & -20.34 & 0.07 & mag \\ 
CIGALE & SFR & 0.4 & 0.6 & $M_\odot \, \rm yr^{-1}$ \\ 
CIGALE & $E(B-V)$ & 0.15 & 0.08 & mag \\ 
pPXF & SFR$_\beta$ & 0.57 & 0.17 &  $M_\odot \, \rm yr^{-1}$ \\ 
\hline 
\end{tabular} 
\end{minipage} 
\end{table}

\clearpage
\begin{table}
\centering
\caption{{\bf 
Physical properties of the foreground galaxy 
\fgname\ 
derived from CIGALE and nebular line analysis
\label{tab:fgprop}}
}
\begin{minipage}{270mm} 
\begin{tabular}{ccccc}
\hline 
Package & Parameter & Value & Uncertainty & Unit 
\\ 
\hline 
CIGALE & AGN Fraction & 0.1 & 0.2 & \\ 
CIGALE & Metallicity & 0.017 & 0.016 & \\ 
CIGALE & Rest-frame $u-r$ & 2.3 & 0.1 & mag \\ 
CIGALE & Rest-frame $M_r$ & -21.98 & 0.05 & mag \\ 
pPXF & $E(B-V)$ & 0.07 & 0.20 & mag \\ 
pPXF & [N{\sc ii}]/H$\alpha$ & 0.28 & 0.03 & log10 \\ 
pPXF & [O{\sc iii}]/H$\beta$ & 1.10 & 0.05 & log10 \\ 
\hline 
\end{tabular} 
\end{minipage} 
\end{table}

\clearpage
\begin{table}
\centering
\caption{{\bf 
Emission line fluxes and luminosities for host galaxy
and foreground galaxy \fgname\ of FRB~181112
measured from the FORS spectroscopy.}
The luminosity values include
an extinction correction. \label{tab:lines}}
\begin{minipage}{270mm} 
\begin{tabular}{cccccc}
\hline 
Galaxy & Line & Flux & Flux Uncertainty & Luminosity$^a$ & Lum. Uncertainty \\ 
&& ($10^{-17}$ erg/s/cm$^2$) & ($10^{-17}$ erg/s/cm$^2$) 
& ($10^{40}$ erg/s) & ($10^{40}$ erg/s) 
\\ 
\hline 
Host& H $\alpha$ & 6.4 & 0.8 & 5.8 & 0.7\\ 
Host& H $\beta$ & 2.9 & 0.2 & 2.6 & 0.2\\ 
Host& H $\gamma$ & 0.7 & 0.2 & 0.6 & 0.2\\ 
Host& {[NII]  6584} & 4.9 & 1.0 & 4.4 & 0.9\\ 
Host& {[OIII]  5007} & 5.4 & 0.3 & 4.9 & 0.3\\ 
Foreground& H $\alpha$ & 5.5 & 0.3 & 3.4 & 0.2\\ 
Foreground& H $\beta$ & 1.8 & 0.2 & 1.2 & 0.1\\ 
Foreground& {[NII]  6584} & 10.5 & 0.5 & 6.5 & 0.3\\ 
Foreground& {[OIII]  5007} & 23.0 & 0.3 & 15.5 & 0.2\\ 
\hline 
\end{tabular} 
\end{minipage} 
\end{table}

\clearpage

\begin{table}
\centering
\caption{{\bf GALFIT S\'{e}rsic profile fits\label{tab:sersic}}}
\begin{minipage}{270mm} 
\begin{tabular}{ccccc}
\hline 
Galaxy & S\'{e}rsic index & Index fit uncertainty & Half-light radius& Half-light radius uncertainty \\ 
 & & & (arcsec) & (arcsec)\\ 
\\ 
\hline 
Host & 1.9 & 0.4 & 0.64 & 0.04\\ 
Foreground & 2.48 & 0.08 & 0.760 & 0.008\\ 
\hline 
\end{tabular} 
\end{minipage}
\end{table} 

\end{document}